\documentclass[useAMS,usenatbib]{mn2e}
\usepackage{multirow}
\usepackage{multicol}
\usepackage{graphicx}
\usepackage{color}
\usepackage{graphicx}
\usepackage{amssymb}
\usepackage{amsmath}
\usepackage{subfigure}
\usepackage{rotating}
\usepackage{longtable}
\usepackage{lscape}

\title[Tidal debris around Galactic globular clusters]{A search for stellar tidal debris of defunct dwarf galaxies around globular clusters in the inner Galactic halo}
\author[Carballo-Bello et al.]
{Julio A. Carballo-Bello$^{1,2,3}$
\thanks{E-mail: jcarball@das.uchile.es},
Antonio Sollima$^{4}$, David Mart\'inez-Delgado$^{5}$,  
\newauthor
Berenice Pila-D\'iez$^{6}$, Ryan Leaman$^{2,3}$,  J\"urgen Fliri$^{2,3}$,  Ricardo R. Mu\~noz$^{1}$
\newauthor
 and Jes\'us M. Corral-Santana$^{7,2,3}$ 
\\
\\
$^{1}$Departamento de Astronom\'ia, Universidad de Chile, Camino El Observatorio 1515, Las Condes, Santiago, Chile \\
$^{2}$Instituto de Astrof\'isica de Canarias (IAC), V\'ia L\'actea s/n, La Laguna E-38205, S/C de Tenerife, Spain\\
$^{3}$Departamento de Astrof\'isica, Universidad de La Laguna, La Laguna E-38205, S/C de Tenerife, Spain\\
$^{4}$INAF - Osservatorio Astronomico di Bologna, via Ranzani 1, I-40127 Bologna, Italy\\
$^{5}$Astronomisches Rechen-Institut, Zentrum f\"ur Astronomie der Universit\"at Heidelberg,  M{\"o}nchhofstr. 12-14, 69120 Heidelberg, Germany\\
$^{6}$Leiden Observatory, Leiden University, Oort Building, Niels Bohrweg 2, NL-2333 CA Leiden\\
$^{7}$Instituto de Astrof\'isica - Pontificia Universidad Cat\'olica de Chile, Av. Vicu\~na-Mackenna 4860, Macul 7820436, Santiago, Chile\\
}

\usepackage{natbib}
\newcommand{\figref}{Figure~\ref}
\newcommand{\tabref}{Table~\ref}
\newcommand{\equref}{Equation~\ref}
\newcommand{\secref}{Section~\ref}

\begin{document}

\pagerange{\pageref{firstpage}--\pageref{lastpage}} \pubyear{2014}

\maketitle

\label{firstpage}

\begin{abstract}

In the hierarchical formation scenario in which the outer halo of the Milky Way 
is the result of the continuous accretion of low-mass galaxies, a fraction of 
the Galactic globular cluster system might have originated in and been accreted 
with already extinct dwarf galaxies. In this context, we expect that the 
remnants of these progenitor galaxies might be still populating the 
surroundings of those accreted globulars. In this work, we present wide-field 
photometry of a sample of 23 globular clusters in the Galactocentric distance 
range $10\leq R_{\rm G} \leq 40$\,kpc, which we use to search for remnants of 
their hypothetical progenitor systems. Our deep photometry reveals the presence 
of underlying stellar populations along the line-of-sight of about half of the 
globulars included in our sample. Among the detections lying in the footprint of the Sagittarius tidal stream, which we 
identify via the comparison with its orbit derived from numerical simulations, only 
Whiting\,1 and NGC\,7492 seem to be inmersed in that remnant at a compatible 
heliocentric distance. We also confirm the existence of a subjacent Main-Sequence feature in the 
surroundings of NGC\,1851. A tentative detection of the 
vast Hercules-Aquila cloud is unveiled in the background of NGC\,7006.

\end{abstract}


\section{Introduction}
\label{intro}

The formation of the outer regions of disc galaxies in the context of the currently most accepted cosmological model, namely Lambda Cold Dark Matter \citep[$\Lambda$CDM, ][]{Peebles1974}, took place via the hierarchical accretion of minor stellar systems, similar to the nowadays Galactic satellite dwarf galaxies \citep{Font2011a}. Numerical simulations based on this model and focused in our Galaxy \citep{Bullock2005,Cooper2010,Font2011,Gomez2013} predict that the Galactic halo might be populated by stellar remnants, vestiges of these accretion events. An important observational effort has been made to validate this theoretical work by detecting stellar tidal streams in the halo of the Milky Way. 

The first satellite dwarf galaxy discovered that is currently in the process of being accreted is the Sagittarius dwarf spheroidal \citep[Sgr dSph; e.g. ][]{Ibata1994,Bonifacio2004,Bellazzini2006a,Siegel2007} which is following an almost polar orbit around the Galaxy. The destruction of this minor system has generated the largest and most complex halo substructure observed so far \citep{Majewski2003,MartinezDelgado2004,Belokurov2006,Koposov2012}, which has allowed for an investigation of the mass distribution - potential - of the Milky Way by reconstructing its orbit from diverse spectroscopic and photometric datasets \citep[e.g. ][]{Law2010a,Penarrubia2010}. However, there are still significant aspects of this substructure  pending for a satisfactory explanation, like the existence of a bifurcation into two parallel streams on its northern section \citep{Fellhauer2006,Penarrubia2011} .

Far from being the only detected tidal debris, wide-sky surveys as the Sloan 
Digital Sky Survey and the Two Micron All Sky Survey \citep[SDSS and 2MASS 
respectively; ][]{York2000,Skrutskie2006}, have revealed the existence of 
substructures  such as the Monoceros ring \citep{Newberg2002,Yanny2003,
Rocha-Pinto2003,Conn2005,Conn2007,Juric2008,Sollima2011}, diverse streams such 
as the Orphan \citep{Grillmair2006b,Belokurov2007,Sales2008,Newberg2010}, Aquarius 
\citep{Williams2011}, Cetus \citep{Newberg2009} and Virgo \citep{Duffau2006}, 
and the over-densities of Hercules-Aquila \citep{Belokurov2007a,Simion2014} and 
Virgo \citep{Juric2008,Martinez-Delgado2007,Bonaca2012} as the best-studied 
examples. In addition, minor mergers and faint substructures have been observed 
in spiral galaxies in the Local Universe \citep[e.g.][]{Ibata2001a,Ibata2007,Martinez-Delgado2008,McConnachie2009,Martinez-Delgado2010}, showing that our Galaxy is not unusual in this respect.

The globular cluster (GC) population of a given galaxy contains valuable 
information about the formation process of its host galaxy. Evidence 
for separate populations of GCs in the Milky Way and other galaxies have been 
steadily accumulating, and it is interpreted as evidence that supports the 
hierarchical galaxy formation scenario \citep{Zinn1993,Leaman2013,Tonini2013}.
In their seminal paper \cite{Searle1978} showed that while GCs in the
inner Galactic halo (at distances $<8$\,kpc) show a clear radial abundance
gradient, GCs in the outer halo do not follow this trend. 
In terms of the relation between the horizontal branch (HB) type and 
metallicity found for GCs (and assuming the age as the \emph{second parameter}), 
\cite{Zinn1993} classified globulars into \emph{old} halo and \emph{young} halo 
clusters, where the latter would correspond to the accreted fraction of 
Galactic GCs. Simulations suggest that whereas the outer halo clusters ($R_{\rm G} \ge $ 15\,kpc) were probably 
formed in small fragments subsequently accreted by the Galaxy \citep[with the 
most massive GCs such as Omega Centauri and M\,54 as the possible remnant cores 
of the disrupted progenitor;][]{vanden2004}, 
an inner component of the Milky Way halo (and possibly a fraction of the halo 
GCs) may have formed \emph{in situ} \citep[e.g. ][]{Zolotov2009}. 
A recent analysis of the relative ages for 55 clusters calculated from the
turn-off magnitude \citep{Marin-Franch2009,VandenBerg2013} 
showed that the GCs age-metallicity relation is bifucarted 
into two distinct groups. Interestingly, these studies find that most of the
outer halo GCs belong to the branch  characterized by the steeper age-metallicity relation although GCs 
belonging to both branches cover comparable age ranges.

Among the Milky Way satellites, Fornax and the core of the Sgr dSph host a 
population of 5 and 4 (at least) GCs respectively \citep{Ibata1997,Strader2003}, 
suggesting that accreted low-mass systems might have contributed with their own 
globulars to the Galactic GC system. The fraction of accreted Galactic clusters 
estimated by \cite{Forbes2010} represents 1/4 of the entire Galactic GC system, 
when considering parameters such as age-metallicity relations, retrograde 
orbits and HB morphologies. A higher fraction of $\sim 50\%$ of accreted GCs 
was estimated by \cite{Leaman2013}, which is also consistent with the 
estimated fraction of accreted halo stars for the Galaxy \citep{Zolotov2009, 
Cooper2013}. In this context, we expect part of the Milky Way GC population to 
be associated with some of the tidal streams that populate the outer halo, 
similar to what has been observed in M\,31, where the location of the 
outer GC system coincides with the streams observed around that galaxy 
\citep{Mackey2010,Mackey2013}. If these GCs were formed in subsequently accreted stellar systems, they
might be still surrounded by the tidal streams generated by the 
disruption of their progenitor satellites. 

The possible association of Galactic GCs with the stellar tidal stream of Sgr 
has been extensively considered in the literature using different methods and 
datasets \citep[e.g.][]{Dinescu2000,Bellazzini2002,Martinez-Delgado2002,Palma2002,
Forbes2004,MartinezDelgado2004,Carraro2009,Forbes2010}. \cite{Bellazzini2003} 
found that, among the Galactic globulars in the distance range $10 
\leq R_{\rm G} \leq 40$\,kpc, there are at least 18 GCs compatible both in 
position and kinematics with the orbit proposed for that dSph galaxy by 
\cite{Ibata1997}. More recently, \cite[][hereafter L10]{Law2010} also investigated the association
of 64 Galactic GCs with the Sgr stream as predicted by \cite{Law2010a}.
In that case, 9 GCs were suggested as systems formed in the interior of the Sgr
dSph, latter accreted by the Milky Way.

The search for Galactic GCs associated with the other possible major
accretion event, the Monoceros ring, has been 
complicated by the uncertainty about the origin and dynamical history of that 
stellar structure and the unknown location of its tentative progenitor galaxy. 
Different formation scenarios have been proposed for the stellar ring, from the 
accretion by the Milky Way of a dwarf companion system 
\citep{Helmi2003,Martin2004,Martinez-Delgado2005,Penarrubia2005,Sollima2011} to 
the distortion or detection of more distant Galactic components 
\citep{Momany2004,Lopez-Corredoira2006,Momany2006,Hammersley2011}. Regarding 
the hypothetical progenitor accreted dwarf galaxy, the controversial Canis 
Major stellar over-density in the direction $(\ell, b) = (240^\circ, -8^\circ)$ 
at $\sim$ 7\,kpc from the Sun has been proposed as its remnant nucleus 
\citep{Martin2004,Dinescu2005,Martinez-Delgado2005,Bellazzini2006a} but its 
origin has also been the subject of debate during the last years 
\citep{Momany2004,Moitinho2006,Mateu2009}. Nonetheless, several low-latitude 
GCs have been proposed as members of the Monoceros progenitor galaxy GC system 
including NGC\,1851, NGC\,1904, NGC\,4590 and Rup\,106 \citep{Martin2004,Forbes2010}, NGC\,2298 \citep{Crane2003,Frinchaboy2004,Martin2004,Forbes2010} and NGC\,7078 \citep{Martin2004}. 

In this work, we explore the possibility of the presence of stellar remnants of accreted dwarf galaxies around a sample of GCs in the inner Galactic halo, which have been extensively considered as tracers of the hierarchical formation of the Milky Way halo. With that purpose, we present wide-field deep photometry of a statistically significant sample of clusters and of the area surrounding them.

\section{OBSERVATIONS AND DATA REDUCTION}
\label{sample}

\begin{table*}
\small
\begin{centering}
\begin{tabular}{rccrrcccc}
Cluster &  $\ell(^{\circ}$) & \emph{b}($^{\circ}$) & $d_{\odot}$(kpc) & $R_{\rm G}$(kpc) & $r_{\rm t}$ (arcmin) & [Fe/H]  \\
\hline
\hline
\\
Whiting \,1 & 161.2 & -60.7 & 30.1 & 34.5 & 3.2 & -0.70   \\
NGC\,1261  &  270.5 & -52.1 & 16.3 & 18.1 & 10.9 & -1.27 \\
NGC\,1851  &  244.5 & -35.0 & 12.1 & 16.6 & 11.6 & -1.18 \\
NGC\,1904  &  227.2 & -29.3 & 12.9 & 18.8 & 11.3 & -1.60 \\
NGC\,2298  &  245.6 & -16.0 & 10.8 & 15.8 & 10.1 & -1.92 \\
NGC\,4147  &  252.8 & 77.2 & 19.3 & 21.4 & 6.6 & -1.80 \\
Rup\,106   &  300.8 & 11.6 & 21.2 & 18.5 & 9.0 &  -1.68 \\  
NGC\,4590  &  299.6 & 36.0 & 10.3 & 10.2 & 21.4 & -2.23 \\
NGC\,5024  &  332.9 & 79.7 & 17.9 & 18.4 & 18.0 &  -2.10 \\
NGC\,5053  &  335.7 & 78.9 & 17.4 & 17.8 & 13.1 &  -2.27 \\
NGC\,5272  &  42.2 & 78.7 & 10.2 & 12.0 & 25.4 &  -1.50 \\
AM\,4      &  320.3 & 33.5 & 32.2 & 27.8 & 3.3  & -1.30  \\ 
NGC\,5466  &  42.2 & 73.6 & 16.0 & 16.3 & 23.4 &  -1.98 \\
NGC\,5634  &  342.2 & 49.3 & 25.2 & 21.2 & 9.6 &  -1.88 \\
NGC\,5694  &  331.1 & 30.4 & 35.0 & 29.4 & 4.7 &  -1.98   \\
NGC\,5824  &  332.6 & 22.1 & 32.1 & 25.9 & 5.7 &  -1.91    \\
Pal\,5     &  0.8 & 45.9 & 23.2 & 18.6 & 21.1 &  -1.41 \\
NGC\,6229  &  73.6 & 40.3 & 30.5 & 29.8 & 3.8 &  -1.47   \\
Pal\,15    &  18.8 & 24.3 & 45.1 & 38.4 & 5.6 &  -2.07  \\ 
NGC\,6864  &  20.3 & -25.7 & 20.9 & 14.7 & 6.8 &  -1.29 \\
NGC\,7006  &  63.8 & -19.4 & 41.2 & 38.5 & 5.7 &  -1.52  \\  
NGC\,7078  &  65.0 & -27.3 & 10.4 & 10.4 & 17.5 &  -2.37 \\
NGC\,7492  &  53.4 & -63.5 & 26.3 & 25.3 & 9.2  &  -1.78 \\
\\
\hline
\hline
\end{tabular}
\caption{\small{Sample of Galactic GCs: positional data, tidal radii and metallicities \citep{Harris2010,Carballo-Bello2012}.  }}
\label{postab}
\end{centering}
\end{table*}

\subsection{Sample selection and observations}
\label{completeness}

The results presented here are part of a systematic survey of stellar tidal 
debris around GCs of the Galactic halo, based on photometric observations of 
these systems with wide-field cameras at different intermediate-size telescopes 
during the last 10 years. Preliminary results of this survey were presented in 
\cite{Martinez-Delgado2002}, \cite{MartinezDelgado2004} and 
\cite{Carballo-Bello2012}. In this work, we have focused on clusters lying in 
the distance range 10 $\leq R_{\rm G} \leq$ 40\,kpc (only 9 Galactic GCs are found 
beyond that distance), which might include the suggested transition region 
between accreted and \emph{in-situ} formed Galactic stellar halo 
\citep[$R_{\rm G} \sim$ 15--20\,kpc][]{Carollo2007}. To minimize the presence 
of disc stars which could severely affect our photometry, we excluded from the 
initial sample all those clusters at low Galactic latitude 
($|b|\le 20^{\circ}$) with the exception of NGC\,2298 and Rup\,106, globulars 
whose properties suggest an external origin 
\citep{Crane2003,Forbes2010,Dotter2011}. We have also excluded NGC\,6715, 
Terzan\,7, Arp\,2 and Terzan\,8 because their location has already been 
established within the main body of the Sgr dSph 
\citep{Bellazzini2003,Bellazzini2008} and Pal\,12, which has previously been 
associated to its stream \citep{Dinescu2000,Martinez-Delgado2002}. Following 
these criteria, our final sample is composed of 23 of these systems. 
It represents a half of the GCs in the considered Galactocentric distance range 
and the 3/4 of the clusters that match our latitude criteria. 
 
\tabref{postab} includes the position, distance and other relevant information of our GC sample. 
The positions and [Fe/H] estimates are taken from the \cite{Harris2010} catalogue. 
Tidal radii are taken from \cite{Carballo-Bello2012}, where structural 
parameters for these globulars were derived using the same photometric database 
presented in this paper, with the exception of the clusters Whiting\,1, AM\,4, Pal\,15 and NGC\,7006 not included in that work and for which they have been estimated using a similar procedure as the described
in \cite{Carballo-Bello2012}. The radial density profiles for these 4 GCs have been obtained including information for the inner region of the clusters from the literature \citep{Harris1991,Trager1995,Carraro2007,Carraro2009}.

\begin{table*}
\small
\begin{centering}
\begin{tabular}{rrrccccl}
Cluster &   RA (2000) &  DEC (2000) &  t$_{\rm exp}$ $B$ (s)& t$_{\rm exp}$ $R/r$ (s) & seeing('') & Instrument & Obs. run date \\
\hline
\hline
\\
Whiting\,  & 02  :  02  :  56 & -03  :  15  :  10 & 4$\times$900 & 6$\times$600    & 1.2   & WFC & 2010/08/17-19 \\
NGC\,1261  & 03  :  13  :  41 & -55  :  25  :  28 & 4$\times$900 & 4$\times$600    & 0.8   & WFI & 2009/11/08-12 \\
NGC\,1851  & 05  :  13  :  04 & -39  :  49  :  58 & 3$\times$900 & 6$\times$600    & 0.6   & WFI & 2005/07 ($s$) \\
           & 05  :  15  :  02 & -40  :  11  :  57 & 4$\times$900 & 6$\times$600    & 0.8   & WFI & 2010/02/14-19\\
NGC\,1904  & 05  :  23  :  29 & -24  :  19  :  21 & 3$\times$900 & 5$\times$600    & 0.5   & WFI & 2005/07 ($s$) \\
           & 05  :  25  :  29 & -24  :  19  :  19 & 3$\times$900 & 5$\times$600    & 0.5   & WFI & 2005/07 ($s$) \\
NGC\,2298  & 06  :  50  :  05 & -35  :  50  :  47 & 4$\times$900 & 4$\times$600    & 0.9   & WFI & 2009/02/19-22 \\
NGC\,4147  & 12  :  09  :  40 & +18  :  20  :  03 & 4$\times$600 & 4$\times$600    & 0.8   & WFC & 2002/05/15-17 \\
Rup\,106   & 12  :  38  :  48 & -51  :  12  :  36 & 4$\times$900 & 6$\times$600    & 0.9   & WFI & 2009/02/19-22 \\
NGC\,4590  & 12  :  38  :  36 & -26  :  31  :  45 & 4$\times$900 & 6$\times$600    & 0.8   & WFI & 2010/02/14-19 \\
NGC\,5024  & 13  :  12  :  30 & +17  :  49  :  59 & 3$\times$900 & 3$\times$600    & 0.7   & WFC & 2002/05/15-17 \\
NGC\,5053  & 13  :  16  :  01 & +17  :  21  :  51 & 4$\times$900 & 6$\times$600    & 0.6   & WFC & 2010/06/11-13 \\
NGC\,5272  & 13  :  41  :  20 & +28  :  45  :  32 & 2$\times$900 & 3$\times$600    & 1.1   & WFC & 2010/05/18 ($s$) \\
           & 13  :  42  :  11 & +28  :  57  :  32 & 3$\times$900 & 6$\times$600    & 0.7   & WFC & 2010/06/11-13 \\
AM\,4      & 13  :  56  :  06 & -27  :  19  :  18 & 6$\times$900 & 6$\times$600    & 0.9   & WFI & 2009/02/19-22 \\
NGC\,5466  & 14  :  03  :  51 & +28  :  16  :  58 & 4$\times$900 & 4$\times$600    & 0.9   & WFC & 2008/05/11-12 \\
NGC\,5634  & 14  :  28  :  45 & -05  :  45  :  21 & 4$\times$900 & 6$\times$600    & 0.9   & WFI & 2010/02/14-19 \\
NGC\,5694  & 14  :  38  :  40 & -26  :  21  :  25 & 4$\times$900 & 5$\times$600    & 0.8   & WFI & 2010/02/14-19 \\
NGC\,5824  & 15  :  04  :  14 & -32  :  53  :  15 & 4$\times$900 & 6$\times$600    & 1.0   & WFI & 2010/05/15-19 \\
Pal\,5     & 15  :  15  :  41 & -00  :  06  :  48 & 2$\times$1000 & 3$\times$900   & 0.9   & WFC & 2001/06/20-27 \\
NGC\,6229  & 16  :  46  :  25 & +47  :  20  :  06 & 3$\times$900  & 5$\times$600   & 1.2   & WFC & 2010/08/17-19  \\
Pal\,15    & 16  :  59  :  36 & -00  :  24  :  45 & 4$\times$900  & 6$\times$600   & 0.9   & WFI & 2010/05/15-19 \\
NGC\,6864  & 20  :  05  :  46 & -21  :  41  :  30 & 3$\times$900 & 6$\times$600    & 0.5   & WFI & 2010/05/15-19 \\
NGC\,7006  & 21  :  01  :  29 & +16  :  11  :  15 & 4$\times$900 & 4$\times$600    & 1.0   & WFC & 2001/06/22-28 \\
           & 21  :  08  :  48 & +18  :  24  :  51 & 3$\times$900 & 4$\times$600    & 0.9   & WFC & 2001/06/22-28 \\
NGC\,7078  & 21  :  29  :  36 & +12  :  09  :  00 & 3$\times$900 & 6$\times$600    & 1.0   & WFC & 2010/06/11-13 \\
           & 21  :  29  :  58 & +12  :  40  :  00 & 4$\times$900 & 6$\times$600    & 0.8   & WFC & 2010/06/11-13 \\
NGC\,7492  & 23  :  09  :  16 & -15  :  49  :  14 & 4$\times$900 & 5$\times$600    & 0.9   & WFI & 2009/11/08-12\\
\\
\hline
\hline
\end{tabular}
\caption{\small{Coordinates, exposure times, mean seeing and dates of the observational campaigns in which the GCs were observed. Service mode observations are denoted by $(s)$.}}
\label{obstab}
\end{centering}
\end{table*}

Our survey strategy was based on obtaining  deep photometric observations in a 
wide field of view (FOV) around the clusters, which allows us to explore for 
the first time their external regions, poorly represented in shallower 
photometric data. In this case, the main tracers of the tidal debris of these 
possible progenitor systems are main-sequence (MS) stars 2--3 magnitudes fainter than the MS turn-off (TO)
 of the old stellar population.  Given the low levels of surface-brightness for known tidal streams \citep[$\mu_{\rm V} > $ 30\,mag\,arcsec$^{-2}$,][]{Martinez-Delgado2001,Majewski2003}, very deep 
colour-magnitude diagrams (CMDs) are needed to get enough statistic of MS-TO stars
in the explored area. In addition, good seeing conditions  are essential to  undertake a reliable decontamination of background galaxies in the CMD, which would otherwise affect the detection of a  MS feature associated to an underlying stellar population in the blue region of the diagram at fainter magnitudes (e.g. see Figures \ref{methodology1} and \ref{methodology2}). 

 Observations have been performed using the Wide Field Camera (WFC) mounted
 at the Isaac Newton telescope (INT), 
 established at El Roque de los Muchachos 
 Observatory on the island of La Palma (Canary Islands) and the Wide Field 
 Imager (WFI) at the MPG/ESO2.2\,m 
 telescope, at the La Silla Observatory (Chile). 
 The WFC provides, with 4 CCDs with a pixel size of 0.333\,arcsec\,pixel$^{-1}$, 
 a total FOV of 34\,arcmin $\times$ 34\,arcmin. The WFI provides a similar FOV 
 of 34\,arcmin $\times$ 33\,arcmin covered by 8 identical CCDs. A summary of 
 the observations is shown in \tabref{obstab}, including the coordinates of 
 each of the pointings. The typical exposure times were 4$\times$900\,s in the 
 $B$ band and  6$\times$600\,s in $R$. The typical seeing was $FWHM<1\arcsec$.
 Daily sky-flats and bias were obtained and used for 
 bias and flat-field correction by means of reduction routines based on \textsc{iraf} standard tasks. A set of \cite{Landolt1992} 
 standard stars were observed during the runs, at different airmass ranges to allow a precise calibration of the final photometric catalogs.

\subsection{Photometry and completeness test}
\label{completeness}

PSF photometry was obtained using \textsc{daophot ii/allstar} \citep{Stetson1987} Our final catalogs only contains objects with $|$SHARP$| \leq$ 0.4, reducing the pollution in the CMDs by background galaxies and allowing us to detect the MS of the tentative underlying streams in the region of the diagram dominated by these non-stellar objects. The aperture correction of our magnitudes were 
performed using bright stellar-shaped objects in the outer regions of the field, far 
from the GC, with $\sigma <$ 0.1. With these criteria, we had a good sample of 
bright stars to compare the PSF fitting from \textsc{allstar} with the  
aperture photometry obtained with \textsc{daophot\,ii/phot}. 
The typical corrections are below  0.2\,mag. To estimate the magnitude 
of our stars outside the atmosphere, we used the extinction coefficients 
computed for each observatory: $A_{\rm B}= 0.22$ and 
$A_{\rm r}= 0.07$\,magnitudes per airmass unit for the Roque de los Muchachos 
Observatory and $A_{\rm B}= 0.19$ and $A_{\rm R}= 0.06$ at La Silla 
Observatory. 

For the calibration of the WFC photometry, we have searched for 
stars present both in our data and in SDSS. Due to the differences between the 
filters used by this survey and those used in our project, we transformed the 
SDSS magnitudes ($ugriz$) to the Johnson-Cousins system ($BR$) using 
\cite{Chonis2008} equations (for stars in the color range 0.08 
$< r - i <$ 0.5 and 0.2 $< g - r <$ 1.4). The brightest subsample of common stars (20--30 stars per chip) was used to obtain a correction factor to apply to our objects and that 
also accounts for the differences between the photometric systems. No significant color trends have been noticed in the comparison between B
and R magnitudes in the WFC and Johnson-Cousin photometric system. For
this reason we applied only a systematic shift. Mean values for these corrections are found to be $C_{\rm B}$ = 25.10$\pm$ 0.08  
and $C_{\rm R}$ =  25.72$\pm$ 0.09. 
For the WFI data, we derived transformations from the 
comparison of the instrumental results for the \cite{Landolt1992} standard 
stars observed and their calibrated magnitudes. The mean values for the transformation coefficients are:

\begin{equation}
 B_{Lan} - B_{inst} = 25.09\,(\,\pm\, 0.09\,) + 0.19\,(\,\pm\, 0.23\,)\,(B - R)_{inst}
\label{landoltb}
\end{equation}

\begin{equation}
 R_{Lan} - R_{inst} = 24.57\,(\,\pm\, 0.07\,) -0.02\,(\,\pm\, 0.02\,)\,(B - R)_{inst}
\label{landoltr}
\end{equation}\\

In order to estimate the completeness of our photometric catalogs in the surrounding area of the clusters, we have 
considered separately the furthest chip with respect to the cluster centre. 
We have included in the images synthetic stars with magnitudes in the range $17 
\leq B,R \leq 26$ and color $0.5<B-R<1.5$, randomly distributed throughout the chip. The total number 
of synthetic stars added in each of the frames was designed not to exceed 
15$\%$ of the number of originally observed sources and have been placed in
separated cells to avoid self-crowding. For each of the globulars, we obtained 50 of these altered images and they were processed with \textsc{daophot\,ii} using the same PSF model derived for the observed stars.

  \begin{figure}
     \begin{center}
      \includegraphics[scale=0.83]{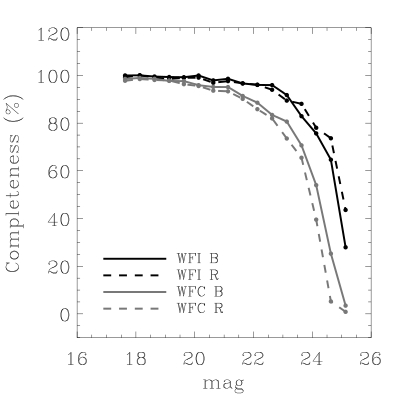}
      \caption[completeness]{{\small Examples of the photometric completeness 
      obtained in this work for the $B$ (solid line) and $R$ (dashed line) 
      bands as a function of the magnitude for WFC (grey) and WFI (black), for the clusters NGC\,6229 and NGC\,5634 respectively. Synthetic stars with magnitudes below 22 in the WFI are completely recovered, while that percentage drops below 80$\%$ at $B,R \sim$ 24. As for the WFC photometry, a completeness of 80$\%$ is derived up to magnitudes $B,R \sim$ 23.5, while it drops to the 60$\%$ at $B,R \sim$ 24. All the magnitudes are in the Johnson-Cousins system.}}
\label{completeness}
     \end{center}
    \end{figure}

We estimated the fraction of synthetic stars recovered by \textsc{allstar} for 
all the images and derived a mean variation of that fraction as a function of 
the magnitude. In \figref{completeness} it is shown the percentage of recovered 
stars for the $B$ and $R$ bands, corresponding to one of the outer chips in the 
mosaics obtained for two clusters with typical exposure times and seeing (see 
\tabref{obstab}), but observed with different instruments. Our results show a 
similar behaviour for both bands but with small differences between the 
instruments. On the one hand, our WFI photometry recovers around the 100$\%$ of 
the synthetic stars up to magnitudes $B,R \sim$ 22 and in that case the 
completeness drops marginally below 80$\%$ at $B,R \sim$ 24. For the WFC, 
considering the same magnitudes ranges defined above, we obtained a 90 
and 60$\%$ of completeness for $B,R \sim$ 22 and 24 respectively. Despite these 
differences regarding the number of recovered sources in both instruments, we 
conclude that, given the depth of our data, our photometry should be able to detect the presence of 
subjacent tidal streams if they are present in the surroundings of these GCs. Hereafter we define $<V> = (B+R)/2$.

\section{METHODOLOGY}
\label{method}

There are several scenarios where one might expect to observe apparent
 tidal debris around Galactic halo GCs: \emph{i)} the GC
 could itself be in the process of tidal disruption  due to the tidal field of the Milky Way 
\citep[e.g. Pal\,5; ][]{Odenkirchen2001,Odenkirchen2003}, \emph{ii)} the tidal debris 
could originate from the disruption of the relic of the galaxy that originally hosted the cluster (as is the case
with the Sgr dwarf GCs and almost certainly those seen apparently embedded in tidal substructures in and around M\,31 \citep{Mackey2013} 
and \emph{iii)} the GC could by chance be superimposed in projection
 against other large-scale halo substructure.

Here, we focus on the detection of debris from a subjacent galaxy remnant.
A complete confirmation of such association between GCs and tidal stream would require
follow-up spectroscopy (e.g. velocities, chemical tagging) for the members
of these tidal debris. With photometry only it is not possible to differentiate between these scenarios, so our
results are a first step to identify those Galactic GCs possibly accreted.

\subsection{Selection of the extra-tidal field of the cluster}

An important issue in this work is to estimate the tidal edge of the cluster 
and separate the possible stellar remnants from the GC stellar content. Tidal 
radii, commonly denoted by $r_{\rm t}$, are key structural parameters in King 
models \citep{King1966} and indicate the distance at 
which the radial density profile reaches the theoretical zero level. It has 
been classically used as the physical edge of a GC and all those stars lying 
beyond this distance have been typically classified as \emph{extra-tidal} 
content. \cite{Carballo-Bello2012} found that when MS stars are
included to derive a more complete radial density profile, the derived 
$r_{\rm t}$ are a $40\%$ bigger on average than those derived from shallower photometry. 
 Moreover, in many cases the overall shape of the
density profile is not well reproduced by King models, especially in the outer
parts of the cluster. This indicates that $r_{\rm t}$ is only a rough estimate of the edge of a
cluster \citep[see also][]{McLaughlin2005} and by assuming it as the separation between 
cluster and fore/background stellar populations, the CMD corresponding to the 
latter might be still populated by cluster members.

\begin{figure*}
     \begin{center}
      \includegraphics[scale=1.05]{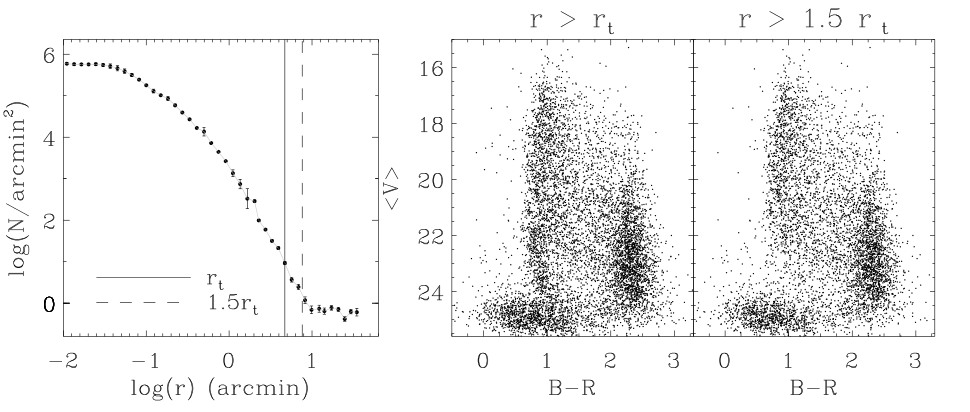}
      \caption[methodology1]{{\small Left: radial density profile derived by \cite{Carballo-Bello2012} for NGC\,5694, where the vertical lines indicate the position of the King tidal radius ($r_{\rm t}$) derived on that work and 1.5 times that distance from the cluster centre. The middle and right panels correspond to the CMDs generated with the stars beyond those distances from the cluster centre. When stars beyond the tidal radius are considered, the CMD shows an over-density associated with the presence of NGC\,5694 stars as expected from the position of $r_{\rm t}$ in relation to the derived radial profile.}}
\label{methodology1}
     \end{center}
    \end{figure*}

\begin{figure*}
     \begin{center}
      \includegraphics[scale=1.8]{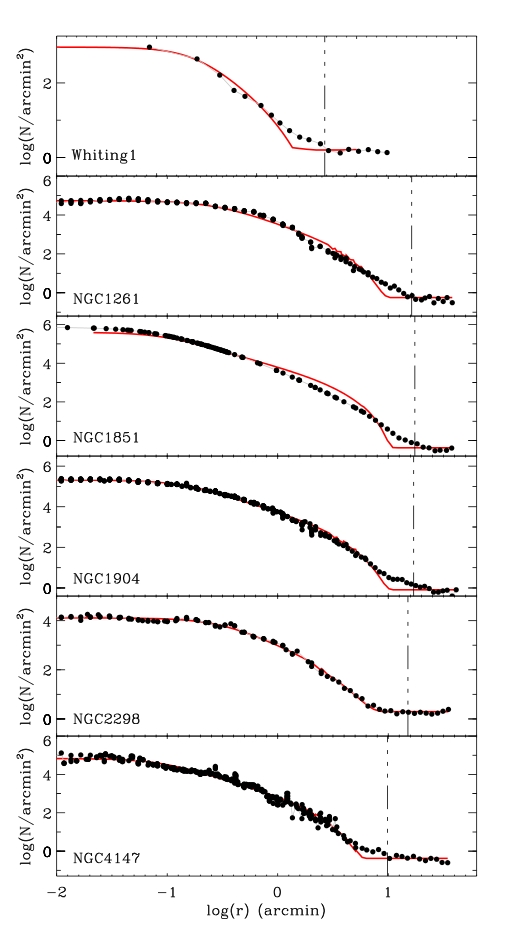}
       \caption{{\small Radial density profiles as derived by  for Whit\,1, NGC\,1261, NGC\,1851, NGC\,1904, NGC\,2298 and NGC\,4147. The vertical line indicate the distances from the cluster centre where the cluster content has been separated from the rest of objects in the photometric catalogues. The red line corresponds with the best King model fitting \citep{Carballo-Bello2012}.}}
\label{radial_detections1}
     \end{center}
    \end{figure*}

\begin{figure*}
     \begin{center}
      \includegraphics[scale=1.8]{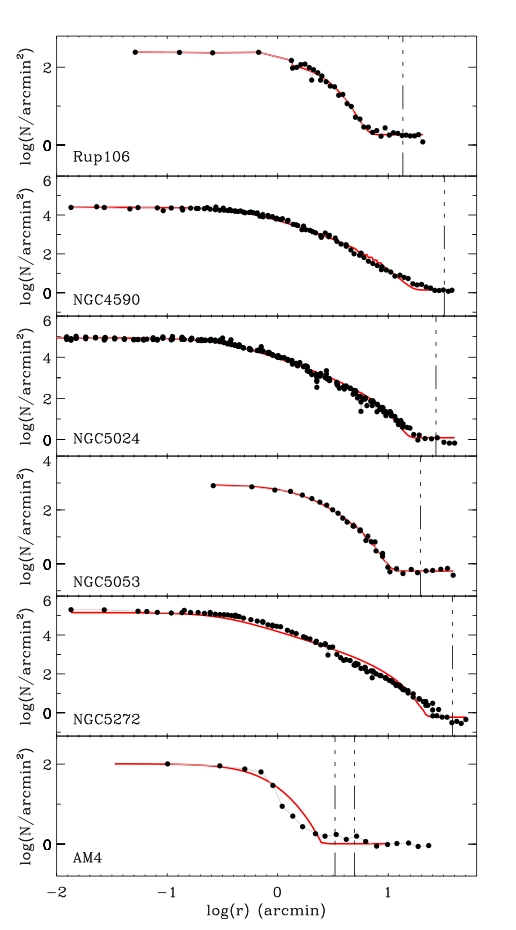}
       \caption{{\small Radial density profiles as derived by  for Rup\,106, NGC\,4590, NGC\,5024, NGC\,5053 and AM\,4. The vertical line indicate the distances from the cluster centre where the cluster content has been separated from the rest of objects in the photometric catalogues. The red line corresponds with the best King model fitting \citep{Carballo-Bello2012}.}}
\label{radial_detections2}
     \end{center}
    \end{figure*}

\begin{figure*}
     \begin{center}
      \includegraphics[scale=1.8]{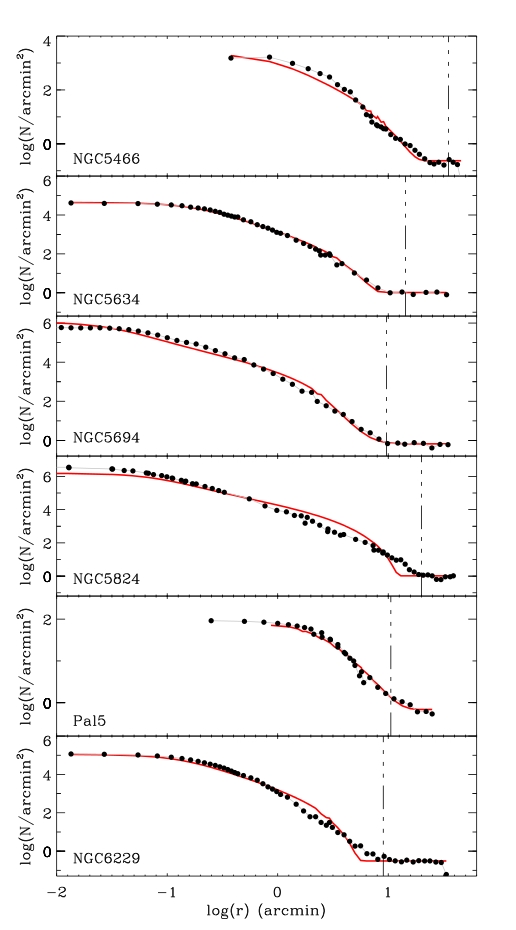}
       \caption{{\small Radial density profiles as derived by  for NGC\,5466, NGC\,5634, NGC\,5694, NGC\,5824, Pal\,5 and NGC\,6229. The vertical line indicate the distances from the cluster centre where the cluster content has been separated from the rest of objects in the photometric catalogues. The red line corresponds with the best King model fitting \citep{Carballo-Bello2012}.}}
\label{radial_detections3}
     \end{center}
    \end{figure*}

\begin{figure*}
     \begin{center}
      \includegraphics[scale=1.8]{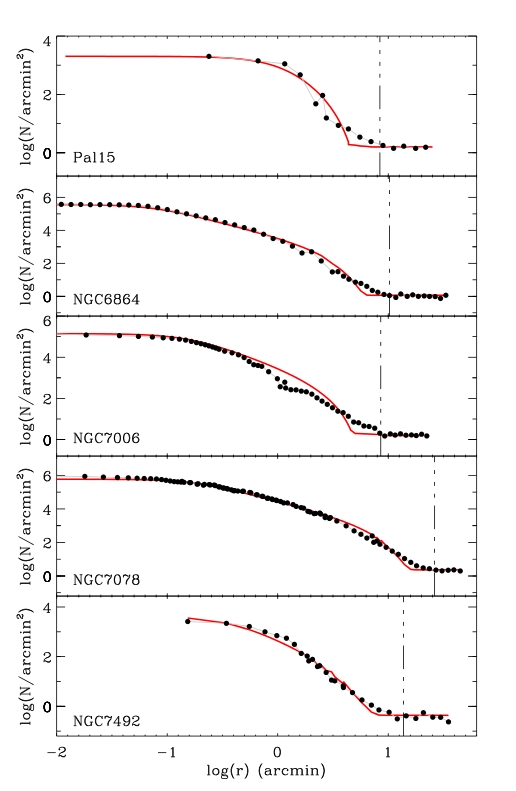}
       \caption{{\small Radial density profiles as derived by  for Pal\,15, NGC\,6864, NGC\,7006, NGC\,7078 and NGC\,7492. The vertical line indicate the distances from the cluster centre where the cluster content has been separated from the rest of objects in the photometric catalogues. The red line corresponds with the best King model fitting \citep{Carballo-Bello2012}.}}
\label{radial_detections4}
     \end{center}
    \end{figure*}

 \figref{methodology1} illustrates the importance of using that selection 
 criteria in the obtention of the CMDs for the fore/background stellar 
 populations. We have generated both the diagrams corresponding to the stars 
 beyond the tidal radius of NGC\,5694 set at $r_{\rm t}= 4.7$\, arcmin and  
 1.5\,$r_{\rm t}$, using in this case the value derived from the profiles 
 obtained in \cite{Carballo-Bello2012}. It is apparent that the King 
 tidal radius lies within the outer part of the GC profile, so the contribution of NGC\,5694 stars becomes important 
 even outside this distance. On the contrary, when the distance at which the radial density 
 remains nearly constant is considered, the over-density associated to the GC 
 content is not present in the diagram. Trying to avoid as much as possible 
 the contamination by GC stars, our criterion for this separation was set on 
 1.5 times the formal King tidal radius $r_{\rm t}$, in most cases 
 this coincides with the distance at which the radial density reaches the background level. We adopted the tidal radii determined in
\cite{Carballo-Bello2012} using the present photometric dataset (listed in \tabref{postab}). For the
clusters Whiting\,1, AM\,4, Pal\,15 and NGC\,7006, not included in the above work, we
determined tidal radii using the same procedure described in
\cite{Carballo-Bello2012}. Hereafter, we define $r_{bg}$ = 1.5\,$r_{\rm t}$. Figures \ref{radial_detections1} to \ref{radial_detections4} show the radial density profiles of our target clusters where the adopted value of $r_{bg}$ is indicated.

Unfortunately, because of the relatively small extra-area and limited angular coverage of our data we are not able to detect any large-scale
gradient and/or asymmetry in the distribution of extra-tidal area.

\subsection{Identification of tidal debris in wide-field photometry with Milky Way synthetic colour-magnitude diagrams}
\label{morfologiaCMD}

Galactic tidal streams are highly dispersed resulting in a low 
surface-brightness structure that generates a modest representation of more 
evolved stars in the CMDs. Thus, we expect that the only feature that may 
indicate the presence of a stream around a GC is the presence of a MS that 
might be coincident with that of the GC if they lie at the same distance. 
However, the same feature could be present if the cluster has developed tidal 
tails because of its interaction 
with the Milky Way.  In the majority of the cases, the MS stars 
from the subjacent tidal remnant are hidden in the CMD due to the combination 
of the contributions of a minor fraction of cluster members, fore/background 
stellar populations from the different Milky Way components (mainly the disc 
and halo) and background galaxies.

The best method to correct from these contribution is to obtain observations of 
adjacent control fields with similar Galactic latitude but several degrees 
away from the GCs, with similar FOV and exposure time than the target fields. However, we 
could not obtain these kind of observations during the observing time granted 
for this project. For this reason, to disentangle  the Milky Way stellar halo 
contribution and to identify any subjacent population, we compared the 
observed diagrams with synthetic CMDs for the same line-of-sight of each 
cluster and for a similar solid angle computed assuming a Milky Way model. In 
this work, we have considered the TRILEGAL \citep{Girardi2005,Vanhollebeke2009} and Besan\c con \citep{Robin2003} Milky Way models, that provides public available webpage scripts to compute simulated CMDs in selected Galactic fields.

\begin{figure}
     \begin{center}
      \includegraphics[scale=0.95]{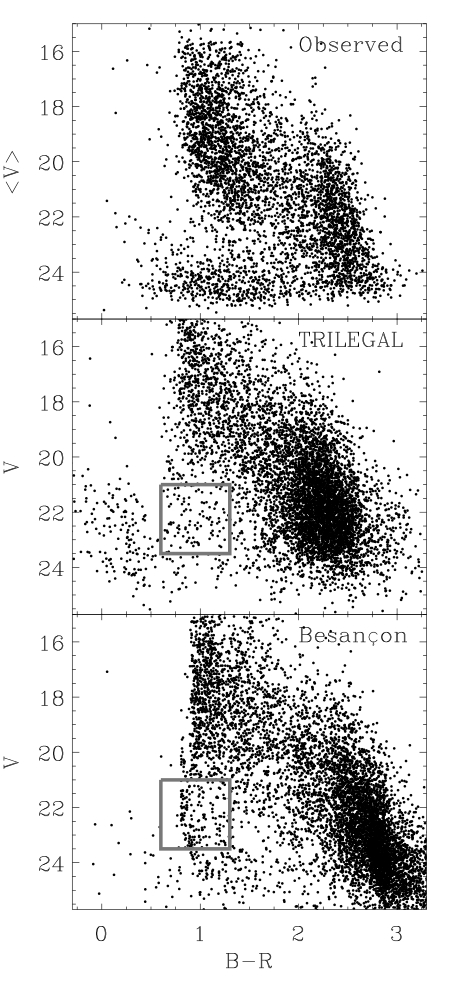}
      \caption[methodology1]{{\small Top: example of CMD obtained for the 
      surroundings of NGC\,2298 for stars beyond $r_{\rm bg}$ from the cluster 
      centre. Middle and bottom panels: CMDs obtained with TRILEGAL and 
      Besan\c con models respectively for a field in the direction of 
      NGC\,2298, with a similar solid angle to that of the observed area around 
      the cluster. The remarkable over-density observed in the bluer region of 
      the observed CMD with $V > 24$ is generated by the presence of background 
      galaxies in the wide-field photometry. In order to compare both synthetic 
      models, we have selected the area in the CMD defined by $0.6 < B-R < 1.5 $ 
      and $21 < V <23.5$ (over-plotted grey rectangle).}}
\label{methodology2}
     \end{center}
    \end{figure}

\figref{methodology2} shows a CMD observed for one of the GCs in our sample 
(NGC\,2298), together with the diagrams obtained with the two models, for the 
same direction in the sky. This comparison allows us to identify the 
over-density of objects in the bluer region of the diagram, around $V \sim 24$, 
as background galaxies, a characteristic feature in wide-field photometry. The 
differences observed between the synthetic CMDs clearly indicate that the 
choice of the Milky Way model for comparison would play a relevant role in the 
detection of Galactic substructures.  In that figure, we have delimited a 
region in the CMD that encompasses the component associated with the Galactic 
halo, in which this study is focused, defined by $0.6< B-R <1.5$ and 
$21 < V < 23.5$. This clearly shows that the Besan\c con model predicts a 
larger number of  stellar halo stars than TRILEGAL, affecting the 
significance of any eventual tidal debris

We have compared the stellar content of TRILEGAL/Besan\c con in that box of the 
CMDs for different sections of the Galactic halo. We have obtained 12 synthetic 
CMDs using both models with an area $\Omega= 0.25$\,deg$^{2}$ for the Galactic 
longitudes $\ell= 0$, $90$ and $180^{\circ}$ and latitudes $b= 25$, $40$, $60$ and 
90$^{\circ}$. The number of predicted halo stars in that box, for all the 
directions in the sky considered, is larger for the Besan\c con results.  For 
$\ell = 180^{\circ}$ and $\ell = 90^{\circ}$ we find a similar behaviour, 
showing that the contribution of halo stars in the TRILEGAL model with 
respect to Besan\c con is considerably lower with $\rm N_{T}/N_{B} \sim$ 
$0.3-0.4$, where $\rm N_{T}$ and $\rm N_{B}$ represent the star counts in that 
box for TRILEGAL and Besan\c con, respectively. These differences might arise 
from the different structural parameters assumed by these models to describe 
the Galactic stellar halo. On the one hand, the TRILEGAL model allows the user 
to select between a $r^{1/4}$ and an oblate $r^{1/4}$ stellar halo 
distribution, where in the latter case the oblateness parameter $q_{\rm h}$ 
remains as free parameter. Instead, in the Besan\c con model, the spheroid component 
is described by a power-law with slope $\alpha =-2.44$ with a fixed value for 
the oblateness set at $q_{\rm h}= 0.76$. 

\cite{Gao2013} has recently studied the ability of these models to reproduce 
Hess diagrams generated from SDSS data in a specific area of the sky. Although 
in their results both models show problems to reproduce the observations, the section
 of the CMD dominated by halo stars - area of interest for this work - 
was more adequately represented by the synthetic diagrams generated by TRILEGAL. 
Given these significant differences in the contribution of halo stars, we will 
continue using as reference both the CMDs generated with TRILEGAL and Besan\c 
con, although new incoming versions of these models, fitting the parameters to 
wide-sky surveys (e.g. Robin et al. in preparation), will have to be taken into 
account in future searches for halo substructures.

To estimate the significance of the detections in our photometry, we have 
compared the observed stellar counts with those computed from the synthetic 
CMDs generated with TRILEGAL for the same line-of-sight and solid angle. The 
input parameters for that model are taken from the optimization obtained by 
\cite{Gao2013} (see Table 3 on that paper). For the Besan\c con model, we 
have used the default parameters. The observed stars considered to derive the 
significance of a subjacent population are those contained between the 
$V$-level of the TO and the level where the CMD is dominated by 
background galaxies, with a difference in colour 0.1$ < \delta (B-R) <$ 0.2 
with respect to the corresponding isochrone (see \secref{isocronas}). Assuming the uncertainty in the number 
counts as $\sigma_{\rm N}$ = $\sqrt{N}$, the significance is given by

\begin{equation}
  S = \frac{(\rm N_{\rm CMD}-\rm N_{\rm model})}{\sqrt{\rm N_{C\rm MD}+ \rm N_{\rm model}}}
\label{definicionS}
\end{equation}

where $N_{\rm CMD}$ is the number of observed stars following the criteria 
described above and $N_{\rm model}$ the TRILEGAL/Besan\c con counts in the same 
area of the synthetic CMD after correcting for completness. In this work, $S$ will indicate the significance of 
the detections with respect to the synthetic model. Given the uncertainties
linked to the performances of the Galactic models in reproducing the real
Galactic field population we defined a conservative treshold for a positive
detection of an underlying stellar population when $S >$5. 

Our ability to detect the presence of stellar substructures with 
surface-brightness comparable to those of Galactic tidal streams is also 
affected by the position of the fields. It is possible to estimate the 
surface-brightness detection limit of our method to detect a Sgr-like stellar 
population that stands out with $S$ = 5 above the Galactic stellar 
populations and as a function of the direction in the sky. We have used the distance-dependent expression proposed by \cite{Bellazzini2006}: 

\begin{equation}
\mu_{V} = -2.5\,{\rm log}(n) + 2.5\,{\rm log}(\Omega) + (m-M)_{0} + K
\label{brilloeq}
\end{equation}

where $n$ is the number of MS stars, $\Omega$ the solid angle observed, 
$(m-M)_{0}$ the distance modulus to the stellar population and $K$ includes 
theoretical elements (see the complete formula in Bellazzini's paper). 
We have calibrated the latter term applying this expression to the 
subjacent {Sgr} population unveiled around Whiting\,1, and using a 
surface-brightness for that portion of the stream 
of $\mu_{\rm V}$ = 30.6\,mag\,arcsec$^{-2}$, measured by \citep{Koposov2012}. 
We define a box in the CMD including all the stars in the subjacent MS to 
determine $K$ - assuming the same heliocentric distance that of Whiting\,1 - 
and used that box in the synthetic CMDs 
used in \secref{morfologiaCMD} to count 
the number of stars predicted by TRILEGAL (after correcting for incompletness). After that, we estimated the 
necessary number of stars in that box to obtain a $S = 5$ detection  above the 
fore/background population using \equref{definicionS} and translate those 
counts into surface-brightness by applying \equref{brilloeq}, assuming the same 
distance modulus of Whiting\,1.

\begin{figure}
     \begin{center}
      \includegraphics[scale=0.6]{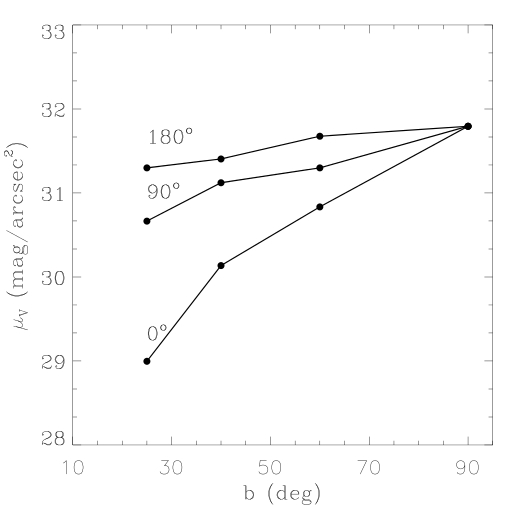}
      \caption[methodology3]{{\small Limiting surface-brightness for three 
      directions in the sky ($\ell=$ 0, 90 and 180$^{\circ}$), defined by the 
      star counts required to obtain a 5$\sigma$ signal above TRILEGAL. As 
      expected, faint substructures as tidal streams will be more easily 
      detectable at higher values of $b$, far from the Galactic stellar 
      components (disc, bulge). }}
\label{methodology3}
     \end{center}
    \end{figure}

\figref{methodology3} shows the limiting surface-brightness (5$\sigma$ detection) as a function of $b$ and for the $\ell$ values considered above. As 
expected, we will be able to detect the presence of fainter halo substructures 
at higher Galactic latitudes, where the halo component becomes less important in the 
obtained CMDs. A tidal stream as the one found around Whiting\,1 would be 
detected in the area $b >$ 80$^{\circ}$ for all $\ell$, when the 
surface-brightness of that structure is as faint as 
31.5 $< \mu_{\rm V} <$ 32\,mag\,arcsec$^{-2}$. The surface-brightness required 
for a tidal stream to be differentiated from the fore/background populations in 
the area around the Galactic centre ($\ell$ = 0$^{\circ}$, $b <$ 40$^{\circ}$) 
is brighter compared to the values obtained for the same stream in the Anticentre direction. These results indicate the areas where faint stellar substructures as the known tidal streams will be more easily detectable.

\subsection{Finding stellar debris  with a cross-correlation algorithm}
\label{cross-correlation}

Given that the detections (and their significances) derived from the synthetic 
CMDs might depend on the selection of the Milky Way model and the input 
parameters, we have also used an alternative approach to look for MS features 
of stellar streams based on the cross-correlation method described in 
\cite{Pila-Diez2014}. This algorithm has been succesfully proven in the case of 
a photometric pencil-beam survey of the Sgr tidal stream using CFHT MegaCam 
deep data but lacking control fields adjacent to the target fields, which is 
the same situation of our GC survey. This method is based on an algorithm that 
takes a CMD as an input and looks across it for the over-density that  best 
matches a template MS population. The template MS is built from the shape of 
an old, metal-poor theoretical isochrone\footnotemark\citep{Marigo2008, 
Girardi10} \footnotetext{This isochrone and all the ones associated to the 
cross-correlation have been retrieved from the Padova Stellar Evolution 
database, available at http://stev.oapd.inaf.it/cmd.} matching the specific 
photometric system of the CMD. The width of this template MS is  tailored to 
the photometric quality of the CMD by accounting for the increase in colour 
error with magnitude of a well defined stellar locus (particularly, the nearby 
M-dwarf stars at $2<B - R<3$).  To each region of the MS template a weight 
based on the distance to the central region of the template is given so that - 
for each step of the cross-correlation - stars placed in the inner part of the template 
have a larger weight than stars at the edges of the template. This 
accounts for possible outliers and statistical contamination.

The algorithm returns two products: the first one is a binned density diagram 
in the colour-magnitude space recording the stellar density contained within 
the template MS for each iteration of the cross-correlation. The second one is 
the MS TO point coordinates (in the colour-magnitude space) for the best match 
(peak of the cross-correlation). We used these binned density diagrams to 
evaluate the quality of the detection by estimating the signal-to-noise of
the cross-correlation procedure and used this last parameter to determine
whether the best match actually represents a real halo feature. We define a positive detection when the S/N
is larger than 5. In all cases with S/N$>$3, we can use the MS TO point magnitude to calculate the distance modulus and the 
heliocentric distance to the substructure (see below). For the templates in 
this work we have used two types of theoretical isochrones: one corresponding 
to the age and metallicity of each GC, which we use on both the corresponding GC 
and on the extratidal field at $r > r_{\rm bg}$ CMDs, and another one corresponding to possible subjacent 
streams, which we only use for the outer-region CMDs.

\subsection{Distances to the underlying populations with isochrone fitting}
\label{isocronas}

\begin{figure}
     \begin{center}
      \includegraphics[scale=0.7]{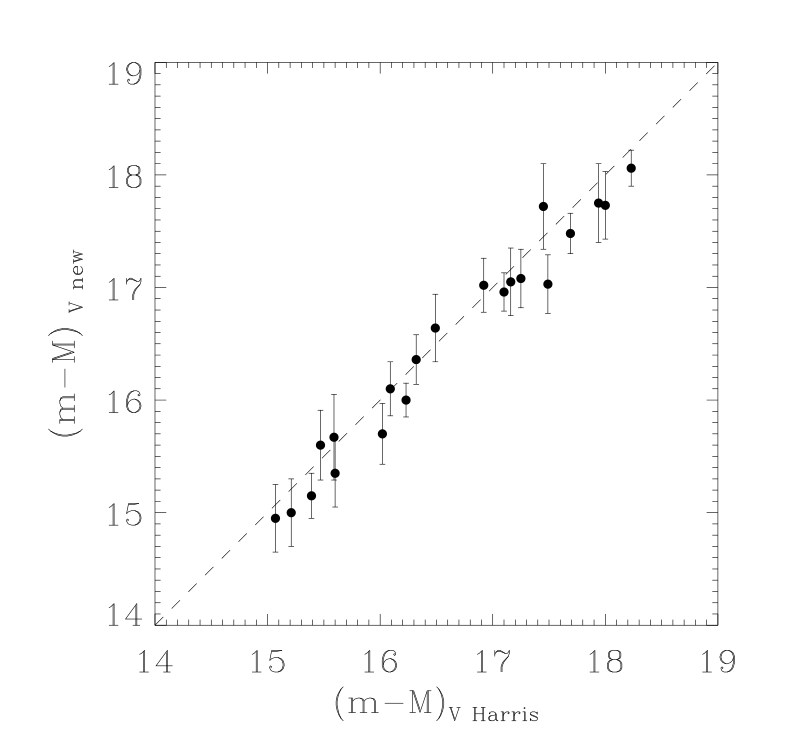}
      \caption[completeness]{{\small Comparison between the \cite{Harris2010} distance modulii and the values derived using the isochrone fitting method described in \secref{isocronas}. $(m-M)_{\rm V}$ and $(m-M)_{\rm V Harris}$ represent the distance modulus obtained for the GCs included in our sample and the one taken from the Harris catalog respectively. The dashed line indicates the 1:1 relation.}}
\label{comparedistances}
     \end{center}
    \end{figure}

Distances to the hypothetical tidal debris are fundamental to conclude if 
they are associated to the GCs or a background, unassociated tidal stream or 
Galactic substructure. Heliocentric distances were derived from the position of 
the MS-feature of the tidal debris in the CMD by fitting a 
reddening-corrected theoretical isochrone. First, the selected isochrone is 
shifted by varying the distance modulus in the range 12 $< (m-M)_{\rm V} <$ 19 
with a step of $\delta (m-M)_{\rm V} =$ 0.2. The $\chi^{2}$ for each position was 
computed taking into account all the stars located in the MS feature (mainly 
populated by the possible stream stars and Galactic halo stellar component). 
The distance modulus value corresponding to the minimum $\chi^{2}$ is then 
selected as initial input for an iterative procedure to obtain a more accurate 
estimate of the position of the isochrone. In this case, we analyzed the 
distance modulus range within a 10$\%$ above and below that value with a 
smaller step $\delta\rm (m-M)_{\rm V}$ = 0.01 ($\sim$150\,pc). This 
fitting method has been tested using the CMD corresponding to the inner regions 
of the GCs, for which we used the isochrones assuming previous estimates for 
their age $t$ and [Fe/H] \citep{Forbes2010,Harris2010}. 
\figref{comparedistances} shows the comparison between the $(m-M)_{\rm V}$ 
obtained for the 23 globulars and those listed in the Harris catalog. 
There is no evidence of systematic deviation from the 1:1 line, this is also 
confirmed by the obtained correlation coefficient ($r$ = 0.97). This test 
confirms that it is possible to obtain the heliocentric distances for stellar 
remnants using our method and, more importantly, there are not systematic 
effects in our photometry, even though is based in 
observing campaigns from two different telescopes spread over ten years. 

Because of the lack of a red-giant branch (RGB) feature associated to the detected tidal debris, it is not possible to obtain insights on their metallicity from our CMDs. This also prevents us from selecting the most suitable isochrone for each case. For this reason, we use only two different cases for our derivation of these distances:  in the case of the remnants possibly associated to the Sgr stream (see \secref{discusionsagitario}), we used an isochrone based on \cite{Siegel2007} results, with an intermediate age of $ t = 6$\,Gyr with [Fe/H] = -0.6. For the over-densities found around other clusters, we assume that they are dominated by an old stellar population similar to those of the typical Milky Way dSph galaxies (i.e. Ursa Minor or Draco). In these cases,  an isochrone with $t = 12$\,Gyr and [Fe/H] = -1.5 is used for the fitting method described above. We have used the theoretical isochrones generated by \cite{Dotter2008} in combination with the Galactic extinction maps from \cite{Schlegel1998} and the extinction coefficients from \cite{Schlafly2011}. Possible effects of spatial variations on the extinction over the FOV are expected to be smalle ($\Delta E(B-V) < 0.02$) even in the clusters at the lowest latitude included in our sample, and have been neglected. The resulting heliocentric distance for the possible tidal remnants detected in this study are given in  \tabref{finalresults}.

In addition to these distance estimates, the MS TO point coordinates for the best match (i.e. peak of the cross-correlation) obtained from the algorithm described in \secref{cross-correlation} were used to calculate the distance modulus and the heliocentric distance to the substructures. The potential age and metallicity gradients of the subjacent streams and their effect on the distance calculation have been included in the uncertainties as discussed in \cite{Pila-Diez2014}. The results of this method are given in \tabref{crosscorrelationresults}.

\section{RESULTS}
\label{results}

\begin{table*}
\footnotesize{
\small
\begin{centering}
\begin{tabular}{rcccllllll}
\hline
  Cluster & A$_{\rm bg}$ (deg$^{2}$)& $d_{\rm GC}$ (kpc) & $d_{\rm bg}$ (kpc) & N$_{\rm CMD}$ & N$_{\rm T}$ &   N$_{\rm B}$ & $S_{\rm T}$ & $S_{\rm B}$ & Note\\
\hline
\hline
\\

NGC\,5634  & 0.21 & 25.7$\pm$ 3.8 & 47.9 $\pm$ 4.7 &  190  & 36.2 & 50.1 &  10.2 $\pm$ 0.8  & 9.0  $\pm$ 0.9    &  Sgr \\

Whiting\,1 & 0.31 & 25.4$\pm$ 3.2 & 26.3 $\pm$ 3.3 &  181  & 35.2 & 20.3 &  10.0 $\pm$ 0.8  & 11.3 $\pm$ 0.7  & Sgr\\   

NGC\,4147  & 0.23 & 21.6$\pm$ 2.3 & 29.3 $\pm$ 4.9 &  254  & 95.7 & 88.2 &  8.5 $\pm$ 0.9  & 9.0 $\pm$ 0.9    & Sgr \\

NGC\,5053  & 0.17 & 13.2$\pm$ 3.7 & 32.8 $\pm$ 5.3 &  86   & 10.0 & 29.4 &  7.8 $\pm$ 0.7  & 5.2 $\pm$ 0.9     & Sgr \\

Pal\,5     & 0.18 & 25.4$\pm$ 2.9 & 52.8 $\pm$ 5.6 &  244   & 106.7 & 58.5 &  7.3 $\pm$ 0.9  & 10.7 $\pm$ 0.8    & Sgr\\

           &      &               & 27.6 $\pm$ 3.8 &  108  & 24.6 & 34.2 &  7.2 $\pm$ 0.8  & 6.2 $\pm$ 0.9    & Tail\\

NGC\,7006  & 0.52 & 40.9$\pm$ 2.1 & 22.3 $\pm$ 2.5 &  705  & 455.4 & 410.2 & 7.3 $\pm$ 1.0  & 8.8 $\pm$ 1.0    &     Her--Aq/? \\  

NGC\,7492  & 0.22 & 26.3$\pm$ 2.1 & 22.0 $\pm$ 3.8 &  136   & 46.3 & 85.2 &  6.6 $\pm$ 0.9  & 3.4 $\pm$ 1.0     & Sgr \\

NGC\,1851  & 0.36 & 24.7$\pm$ 2.0 & 11.9 $\pm$ 2.0 &  227   & 125.2 & 90.2 &  5.4 $\pm$ 1.0  & 7.7 $\pm$ 0.9     & Mon?/GC Halo \\

NGC\,1261  & 0.21 & 16.6$\pm$ 2.0 & 14.9 $\pm$ 2.6 &  151   & 71.4 & 71.8 &  5.3 $\pm$ 1.0  & 5.3 $\pm$ 1.0    & Tail? \\ 

NGC\,5024  & 0.13 & 18.7$\pm$ 2.0 & 37.7 $\pm$ 5.7 &  48   &  15.7 & 26.0 &  4.1 $\pm$ 0.9  & 2.6 $\pm$ 1.0     & Sgr   \\

NGC\,7078  & 0.20 & 10.0$\pm$ 1.7 & 14.4 $\pm$ 3.7 &  218  & 160.2 & 168.3 & 3.0 $\pm$ 1.0  & 2.5 $\pm$ 1.0    & Her--Aq/? \\

NGC\,1904  & 0.32 & 13.6$\pm$ 2.1 & 15.4 $\pm$ 2.4 &  132  & 98.4 & 94.2 & 2.2 $\pm$ 1.0  & 2.5 $\pm$ 1.0    & ?/GC Halo \\   

NGC\,6229  & 0.24 & 35.0$\pm$ 3.1 & 17.7 $\pm$ 3.2 &  98  & 71.9  & 82.2 & 2.0 $\pm$ 0.9  & 1.2 $\pm$ 0.9    & Her--Aq/?\\ 

\\

\hline
\hline
\end{tabular}
\caption[Results]{\small{Results from the isochrone fitting method described in \secref{isocronas}. Areas observed, distances derived for the GCs and for the subjacent populations are given in the columns labeled with A$_{\rm bg}$, d$_{\rm GC}$ and d$_{\rm bg}$ respectively. Significance of the detections $S_{T}$ and $S_{B}$, with respect to TRILEGAL and Besan\c con respectively, are computed using the number of star counts in the observed CMD ($N_{\rm CMD}$) and the counts for the same area of the diagrams obtained with the model considered here ($N_{\rm T}$ or $N_{\rm B}$). The identity of the subjacent population is suggested in function of the projected position of the sample and the considered model. Clusters ordered by $S_{\rm T}$.}}%
\label{finalresults}
\end{centering}
}
\end{table*}

\begin{figure*}
     \begin{center}
      \includegraphics[scale=0.98]{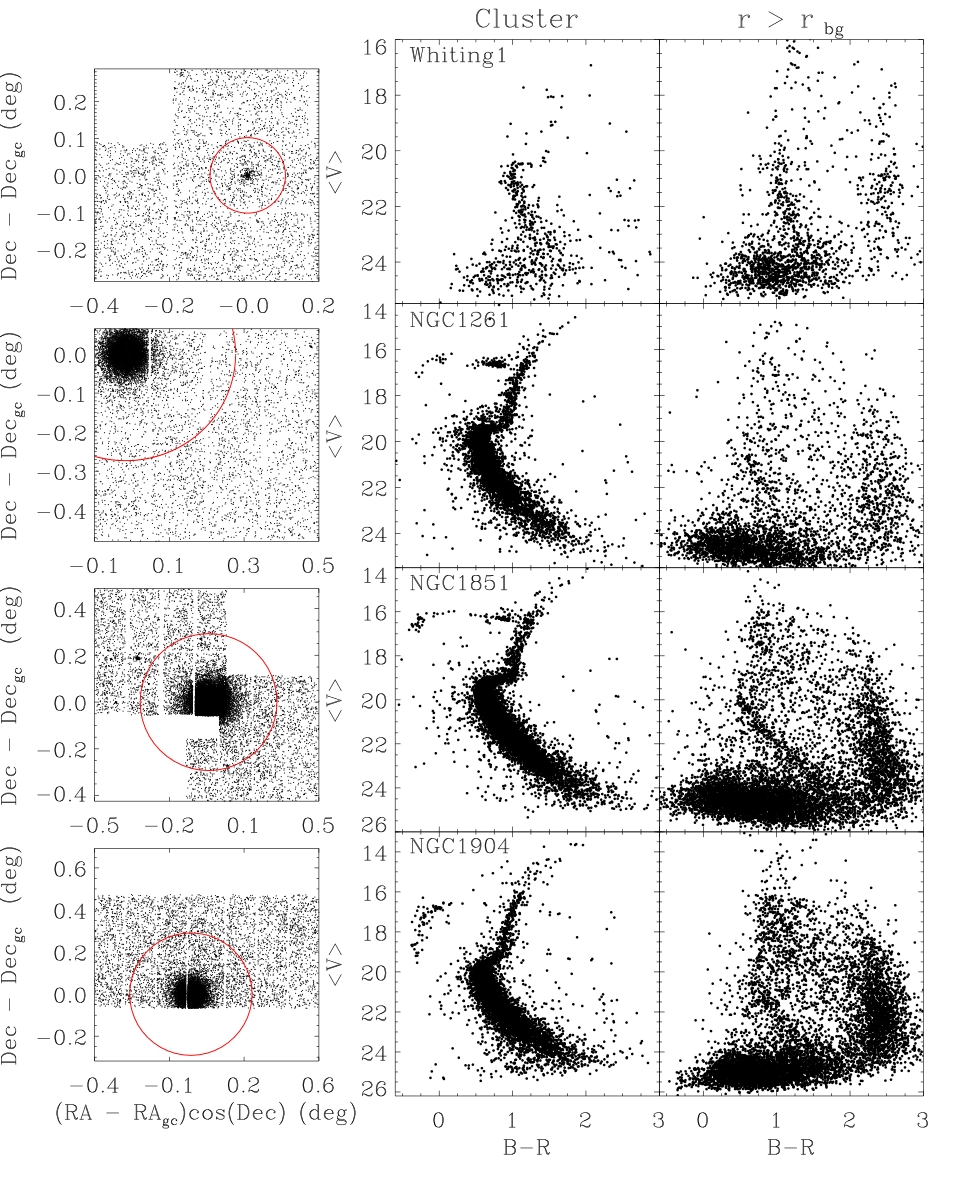}
       \caption[dia1]{{\small CMDs corresponding to the clusters Whiting\,1, 
       NGC\,1261, NGC\,1851 and NGC\,1904 (middle column) and to those objects 
       beyond $r_{\rm bg}$ from the cluster centre (right column). A map 
       showing the distribution of the stars in the catalog with respect to the 
       cluster centre is also included (left), where $r_{\rm bg}$ is indicated by a red line.}}
\label{diagrams1}
     \end{center}
    \end{figure*}

\begin{figure*}
     \begin{center}
      \includegraphics[scale=0.98]{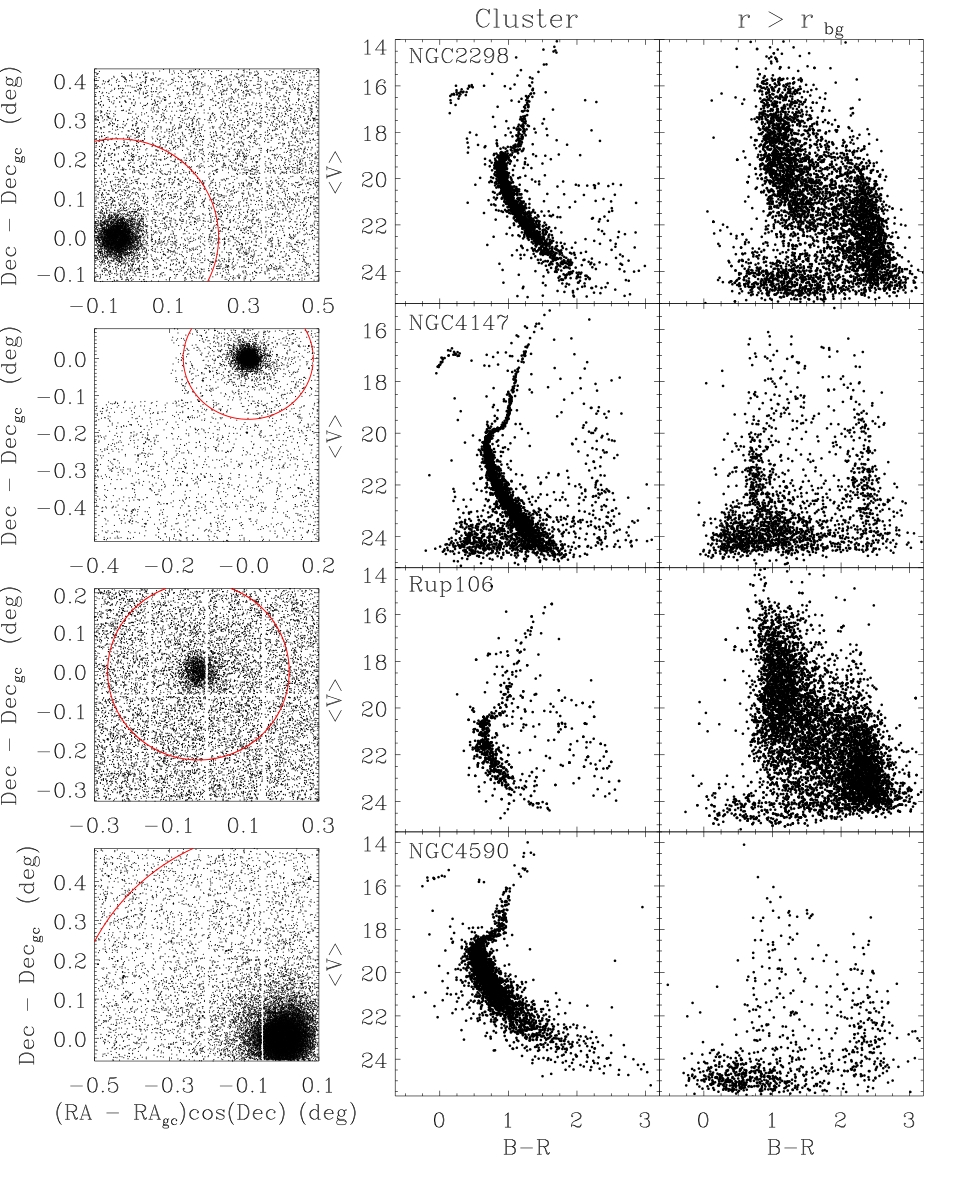}
       \caption[dia2]{{\small CMDs corresponding to the clusters NGC\,2998, NGC\,4147, Rup\,106 and NGC\,4590 (middle column) and to those objects beyond $r_{\rm bg}$ from the cluster centre (right column). A map showing the distribution of the stars in the catalog with respect to the cluster centre is also included (left), where $r_{\rm bg}$ is indicated by a red line.}}
\label{diagrams2}
     \end{center}
    \end{figure*}

\begin{figure*}
     \begin{center}
      \includegraphics[scale=0.98]{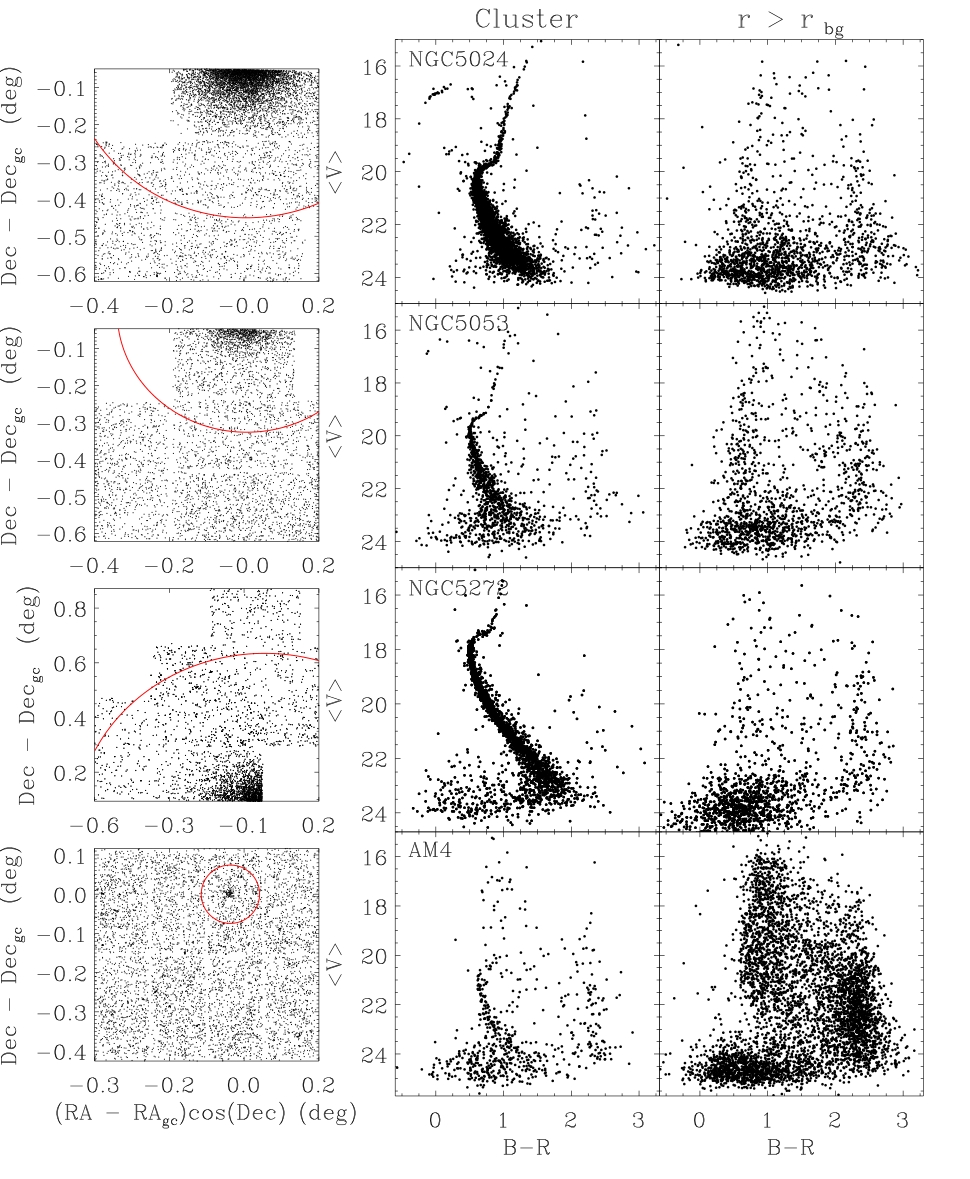}
       \caption[dia3]{{\small  CMDs corresponding to the clusters NGC\,5024, NGC\,5053, NGC\,5272 and AM\,4 (middle column) and to those objects beyond $r_{\rm bg}$ from the cluster centre (right column). A map showing the distribution of the stars in the catalog with respect to the cluster centre is also included (left), where $r_{\rm bg}$ is indicated by a red line.}}
\label{diagrams3}
     \end{center}
    \end{figure*}

\begin{figure*}
     \begin{center}
      \includegraphics[scale=0.98]{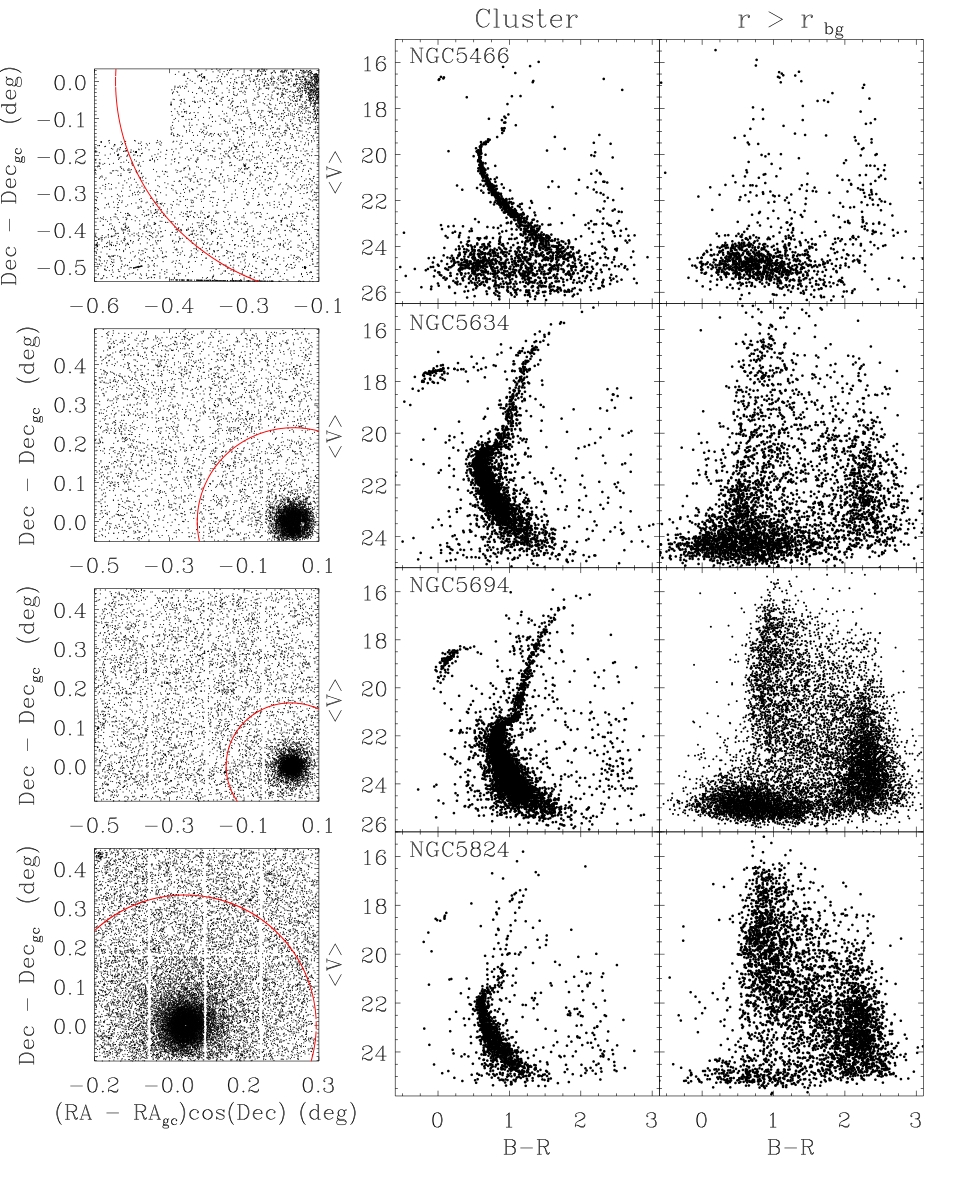}
       \caption[dia4]{{\small  CMDs corresponding to the clusters NGC\,5466, NGC\,5634, NGC\,5694 and NGC\,5824 (middle column) and to those objects beyond $r_{\rm bg}$ from the cluster centre (right column). A map showing the distribution of the stars in the catalog with respect to the cluster centre is also included (left), where $r_{\rm bg}$ is indicated by a red line.}}
\label{diagrams4}
     \end{center}
   \end{figure*}

\begin{figure*}
     \begin{center}
      \includegraphics[scale=0.98]{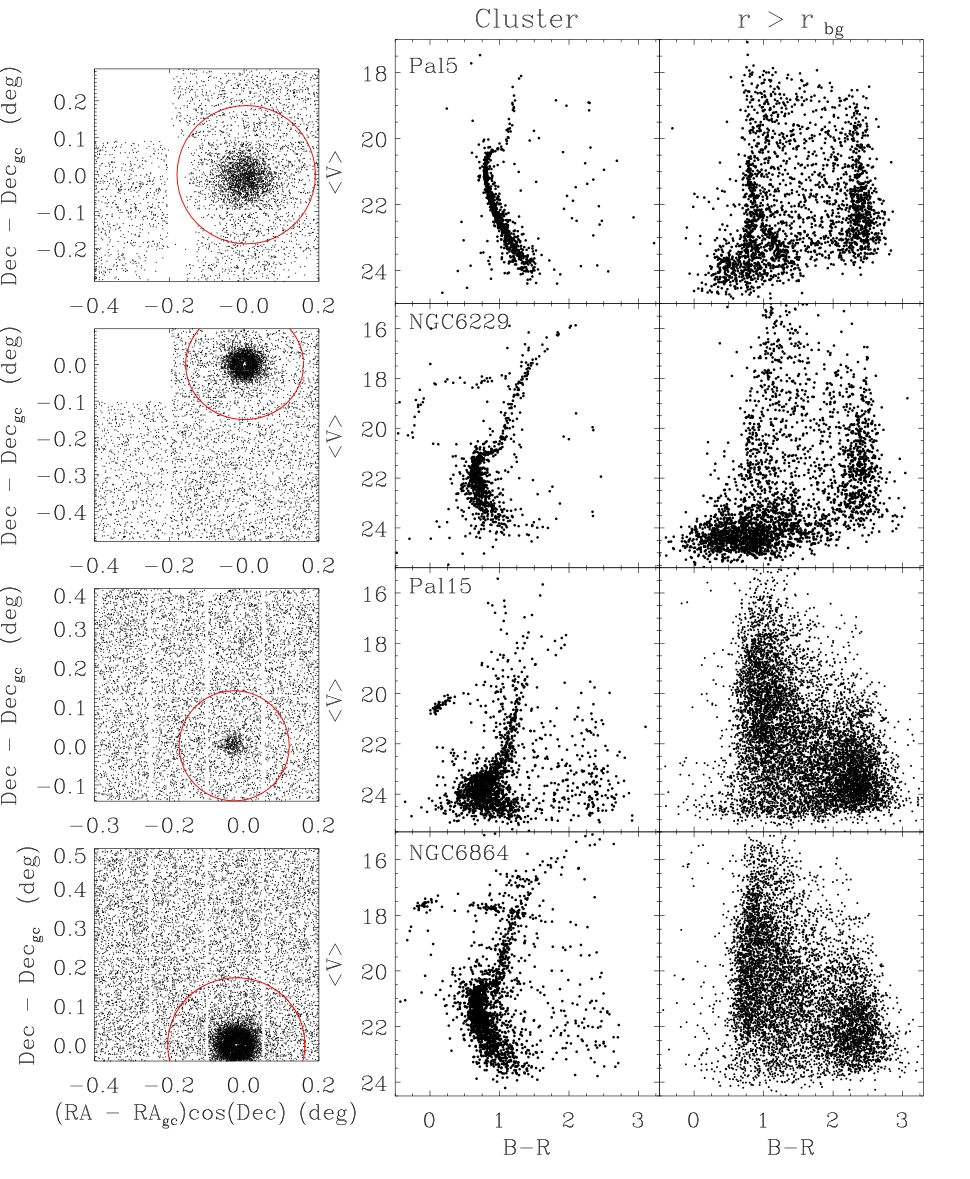}
       \caption[dia5]{{\small  CMDs corresponding to the clusters Pal\,5, NGC\,6229, Pal\,15 and NGC\,6864 (middle column) and to those objects beyond $r_{\rm bg}$ from the cluster centre (right column). A map showing the distribution of the stars in the catalog with respect to the cluster centre is also included (left), where $r_{\rm bg}$ is indicated by a red line.}}
\label{diagrams5}
     \end{center}
    \end{figure*}

\begin{figure*}
     \begin{center}
      \includegraphics[scale=0.98]{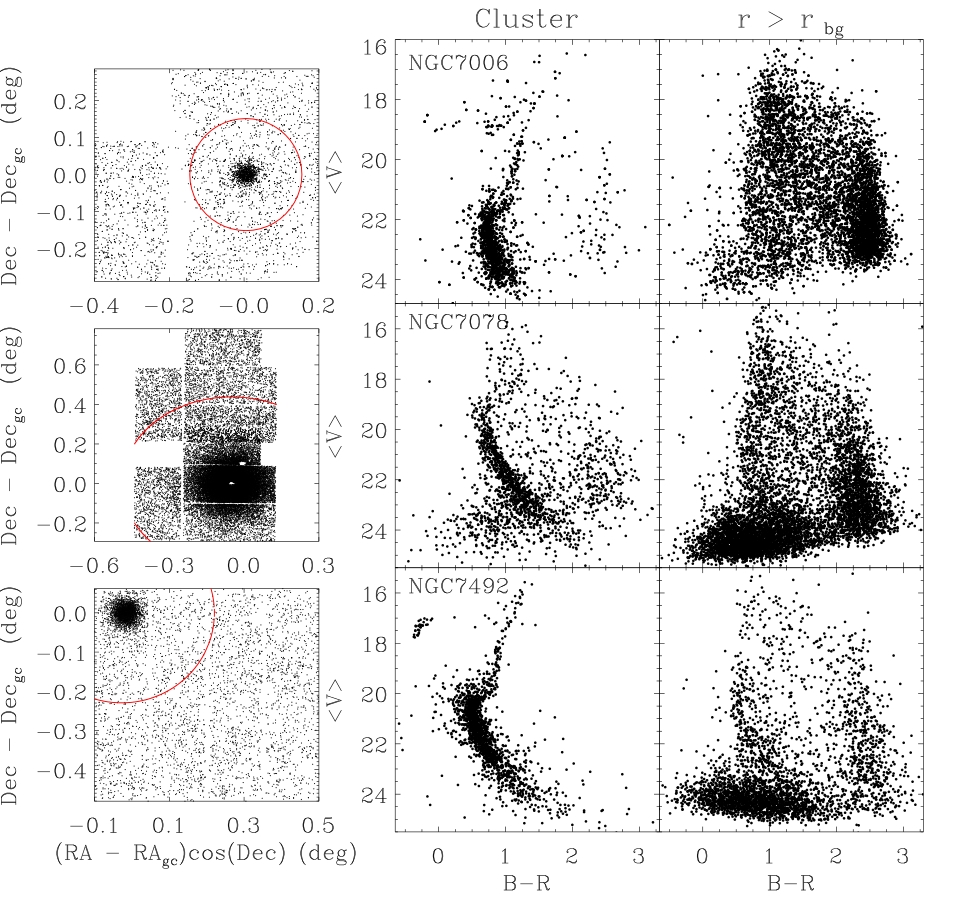}
       \caption[dia6]{{\small  CMDs corresponding to the clusters  NGC\,7006, NGC\,7078 and NGC\,7492 (middle column) and to those objects beyond $r_{\rm bg}$ from the cluster centre (right column). A map showing the distribution of the stars in the catalog with respect to the cluster centre is also included (left). Note that in the case of NGC\,7006, only one of the pointings has been included in that map, where $r_{\rm bg}$ is indicated by a red line.}}
\label{diagrams6}
     \end{center}
    \end{figure*}

Figures from \ref{diagrams1} to \ref{diagrams6} show the calibrated CMDs for 
the GGs in our sample.  For each  cluster, we show the CMD for its central 
region (middle panel) and for those stars situated at a distance beyond 
$r_{\rm bg}$ from the cluster centre (right panel; with the exception of 
Pal\,5, see below). In order to avoid crowding problems, we have included only 
a fraction of the central regions of the cluster confined between an arbitrary 
distance from the centre and the half-mass radius ($r_{\rm h}$), which 
generates the differences in limiting magnitude between the diagrams in some of 
the clusters (e.g. NGC\,6229 and NGC\,4590). The left panels display the 
position of the stellar sources considered in our final photometric catalogs 
with respect to the position of the cluster centre. This provides a good 
reference for the sky area (in degrees; see \tabref{finalresults}) covered 
around each target in this work. The total area observed around each cluster  was 
estimated taking into account the gaps between the chips at both instruments and the position of the cluster 
centre in the field.

The significance of the underlying populations  by means of the 
comparison with a synthetic CMD from the TRILEGAL and Besan\c con Galactic 
models is shown in \tabref{finalresults}. The number of observed stars 
($N_{\rm CMD}$) and the TRILEGAL and Besan\c con counts ($N_{\rm T}$ and 
$N_{\rm B}$ respectively) are used to calculate $S_{\rm T}$ and $S_{\rm B}$ 
using \equref{definicionS}. In \tabref{finalresults} we show the derived 
heliocentric distances for their $r > r_{\rm bg}$ populations. Given that the $S$ values depend clearly on the synthetic Milky Way model and 
the input parameters used (\tabref{finalresults}), our positive detections are 
compared with the results obtained from the application of the 
cross-correlation method to the region defined by $0.0<B-R<1.6$ and 
$18.0< V <24.0$ in the $r > r_{\rm bg}$ CMDs. According to these results, we 
group the clusters in the following categories:

\begin{table*}
 \centering
 \begin{tabular}{lll rc ccc c} 
\hline 
Cluster & Field 	& Group 	&   S/N     & TO$_{\rm V}$ 	&   $(m-M)_{\rm V \ GC-iso}$ 	& $d_{\rm GC-iso}$ (kpc) & $(m-M)_{\rm V \ Sgr-iso}$ &   $d_{\rm Sgr-iso}$ (kpc)\\
\hline
\hline
\\
  am4 		& in 	&   & 5.4 & 21.0 & 17.2 & 28 $\pm$ 2 & 17.2 & 28 $\pm$ 2\\
  am4 		& out 	& A &  &  &  &      &  & \\
  ngc1261 	& in 	&   & 4.4  & 19.8 & 15.9 & 15 $\pm$ 1 & 16.0 & 16 $\pm$ 1\\
  ngc1261 	& out 	& D & 6.9  &  20.0 & 16.1 & 16 $\pm$ 2 & 16.2 & 18 $\pm$ 2\\
  ngc1851 	& in 	&   & 3.9 &  19.4 & 15.5 & 13 $\pm$ 1 & 15.6 & 13 $\pm$ 1\\
  ngc1851 	& out 	& D & 7.3 & 19.0 & 15.1 & 11 $\pm$ 1 & 15.2 & 11 $\pm$ 1\\
  ngc1904 	& in 	&   & 4.0 & 19.8 & 15.9 & 15 $\pm$ 1 & 16.0 & 16 $\pm$ 1\\
  ngc1904 	& out 	& A & 4.4 & 20.2 & 16.3 & 18 $\pm$ 2 & 16.4 & 19 $\pm$ 2\\
  ngc2298 	& in 	&   & 6.4 & 19.5 & 15.5 & 13 $\pm$ 1 & 15.7 & 14 $\pm$ 1\\
  ngc2298 	& out 	& A &  &  &  &  &  &     \\
  ngc4147 	& in 	&   & 6.7 & 20.5 & 16.6 & 21 $\pm$ 1 & 16.7 & 22 $\pm$ 1\\
  ngc4147 	& out 	& D & 5.0 & 21.9 & 18.0 & 41 $\pm$ 6 & 18.1 & 42 $\pm$ 6\\
  ngc4590 	& in 	&   & 4.5 & 19.1 & 15.4 & 12 $\pm$ 1 & 15.3 & 12 $\pm$ 1\\
  ngc4590 	& out 	& A &  &  &  &      &  & \\
  ngc5024 	& in 	&   & 5.4 & 20.4 & 16.5 & 20 $\pm$ 1 & 16.6 & 21 $\pm$ 1\\
  ngc5024 	& out 	& A &  &  &  &  &      & \\
  ngc5053 	& in 	&   & 4.9 & 19.9 & 16.1 & 17 $\pm$ 1 & 16.1 & 17 $\pm$ 1\\
  ngc5053 	& out 	& C &  &  &  &  &     & \\
  ngc5272 	& in 	&   & 6.0 & 18.1 & 14.2 & 6.8 $\pm$ 0.3 & 14.3 & 7.3 $\pm$ 0.3\\
  ngc5272 	& out 	& A &  &  &  &  &      & \\
  ngc5466 	& in 	&   & 5.9 & 19.9 & 15.9 & 15 $\pm$ 1 & 16.1 & 17 $\pm$ 1\\
  ngc5466 	& out 	& A & &  &  &  &      & \\
  ngc5634 	& in 	&   & 4.3 & 21.1 & 17.2 & 28 $\pm$ 1 & 17.3 & 29 $\pm$ 1\\
  ngc5634 	& out 	& D & 6.5 & 22.4 & 18.5 & 51 $\pm$ 9 & 18.6 & 53 $\pm$ 10\\
  ngc5694 	& in 	&   & 5.0 & 22.1 & 18.1 & 42 $\pm$ 2 & 18.3 & 46 $\pm$ 2\\
  ngc5694 	& out 	& A &  &  &  &  &  &     \\
  ngc5824 	& in 	&   & 4.9 & 22.1 & 18.1 & 42 $\pm$ 2 & 18.3 & 46 $\pm$ 2\\
  ngc5824 	& out 	& A &  &  &  &  &  &     \\
  ngc6229 	& in 	&   & 5.3 & 21.5 & 17.5 & 32 $\pm$ 2 & 17.7 & 35 $\pm$ 2\\
  ngc6229 	& out 	& A &  &  &  &  &      & \\
  ngc6864 	& in 	&   & 4.7 & 20.9 & 17.0 & 26 $\pm$ 1 & 17.1 & 27 $\pm$ 1\\
  ngc6864 	& out 	& A &  	&  &  &  &  &     \\
  ngc7006 	& in 	&   & 4.8 & 22.2 & 18.3 & 45 $\pm$ 2 & 18.4 & 48 $\pm$ 2\\
  ngc7006 	& out 	& C &  	&  &  &  &  &     \\
  ngc7078 	& in 	&   & 4.1 & 19.8 & 15.9 & 16 $\pm$ 1 & 16.0 & 16 $\pm$ 1\\
  ngc7078 	& out 	& A &  	&  &  &  &  &     \\
  ngc7492	& in 	&   & 4.7 & 20.5 & 16.6 & 21 $\pm$ 1 & 16.7 & 22 $\pm$ 1\\
  ngc7492 	& out 	& B & 5.4 & 20.2 & 16.3 & 18 $\pm$ 2 & 16.4 & 19 $\pm$ 2\\
  pal15 	& in 	&   &  &   	&  &  &  &     \\
  pal15 	& out 	& A &  	  &  &  &  &  &     \\
  pal5 		& in 	&   & 6.4 & 20.8 & 17.0 & 25 $\pm$ 1 & 17.0 & 25 $\pm$ 1\\
  pal5 		& out 	& D & 5.0 & 22.6 & 18.8 & 58 $\pm$ 6 & 18.8 & 58 $\pm$ 6\\
  pal5 		& out 	& D & 6.1 & 20.8 & 17.0 & 25 $\pm$ 2 & 17.0 & 25 $\pm$ 2\\
  rup106 	& in 	&   & 4.7 & 21.0 & 17.2 & 27 $\pm$ 1 & 17.2 & 28 $\pm$ 1\\
  rup106 	& out 	& A &  &  &  &  &  &     \\
  whit1 	& in 	&   & 5.4 & 20.8 & 17.3 & 29 $\pm$ 3 & 17.0 & 25 $\pm$ 2\\
  whit1 	& out 	& D & 5.2 & 20.5 & 17.0 & 26 $\pm$ 2 & 16.7 & 22 $\pm$ 1\\
\hline
\end{tabular}
\caption[Results_ccorrelation]{\small Cross-correlation results for every field (both inner and outer). For every field we indicate whether there is a detection (D) or not (B), or if the field presents any problem for the method (A and C). For all the inner cases and the outer D cases, we include the cross-correlation MS TO point in the $V$ band. For these fields we also provide the distance modulus and heliocentric distance as derived from two different theoretical isochrones: one representing the stellar population of the nearby GC ($d_{\rm GC-iso}$) and the other one representing that one of the Sgr stream ($d_{\rm Sgr-iso}$).}
\label{crosscorrelationresults}
\end{table*}

\begin{figure*}
   \centering
   \includegraphics[scale=0.3]{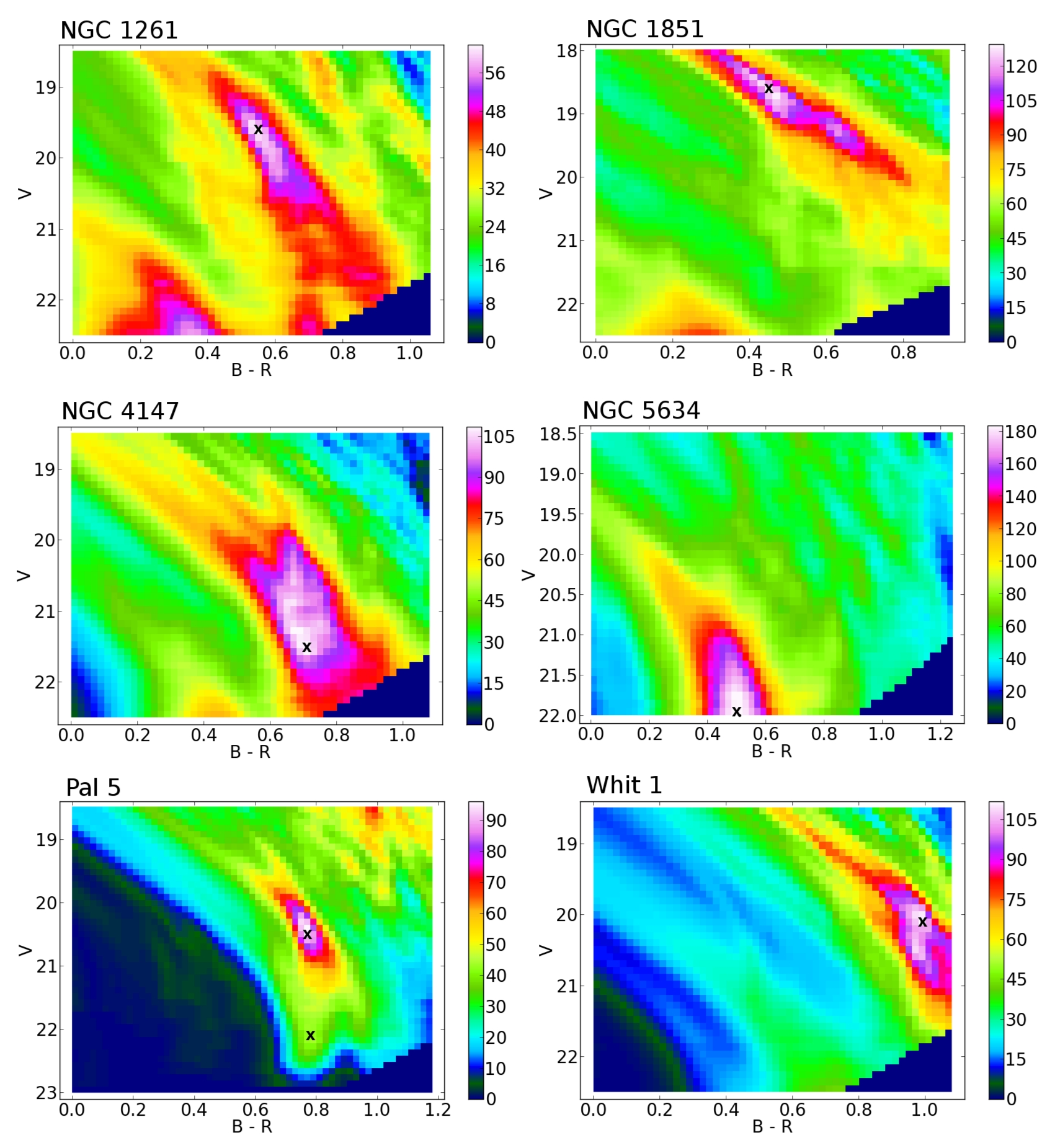}
      \caption{Density diagrams resulting from cross-correlating the CMDs of the outer regions 
               with the MS template. 
               From left to right and top to bottom: NGC1261, NGC1851, NGC4147, NGC5634, 
                Pal\,5 (twice) and Whiting\,1. }
         \label{fig:ccDensityDiagrams}
   \end{figure*}

\begin{itemize}
\item Group A: Clusters for which neither the comparison with Galactic models nor
the cross-correlations return significant detections ($S<5$; $S/N<5$). These 
CMDs correspond to the clusters AM4, NGC1904, NGC\,2298, NGC4590, NGC5024, 
NGC5272, NGC5466, NGC\,5694, NGC\,5824, NGC6229, NGC\,6864, 
NGC\,7078, Pal\,15 and Rup\,106. We refer to this group as "no detections".

\item Group B: Clusters for which an overdensity with $S>5$ is detected with
respect one of the adopted reference
Galactic field models and
the CMD cross-correlation provides a good match with $S/N>5$. The only cluster 
in this group is NGC7492. We refer to this group as "uncertain" detections.

\item Group C: Clusters for which an overdensity with $S>5$ using both reference
Galactic field models is detected but the CMD cross-correlation provides an inconclusive result. The CMDs in this 
group correspond to NGC5053 and NGC7006. We refer to this group as "possible" 
detections.

\item Group D: Clusters for which an overdensity with $S>5$ using both reference
Galactic field models is detected and the 
CMD cross-correlation identifies a distinct MS with $S/N>5$ and pins its TO 
point. The CMDs in this group correspond to NGC\,1261, NGC\,1851, NGC\,4147, 
NGC\,5634, Pal\,5 (twice) and Whithing\,1. We refer to this group as "probable" 
detections. Their density diagrams are shown in \figref{fig:ccDensityDiagrams}.
\end{itemize}

The distance moduli and heliocentric distances to the structures belonging to 
group D are calculated using the cross-correlation algorithm and the two possible isochrones mentioned above 
(either the one from the nearby GC or the one from the Sgr stream). The 
derived distances (\tabref{crosscorrelationresults}) are 
consistent with those obtained using the isochrone fitting method given in Sec. 
3.4, without any evidence of systematic offset or trend. We thus conclude 
that the cross-correlation method independently confirms (within the 
uncertainties) the distance measurements for the GCs classified as group D. 

\section{DISCUSSION}

\subsection{Overdensities associated with the Sagittarius tidal stream}
\label{discusionsagitario}
   \begin{figure*}
     \begin{center}
      \includegraphics[scale=1.8]{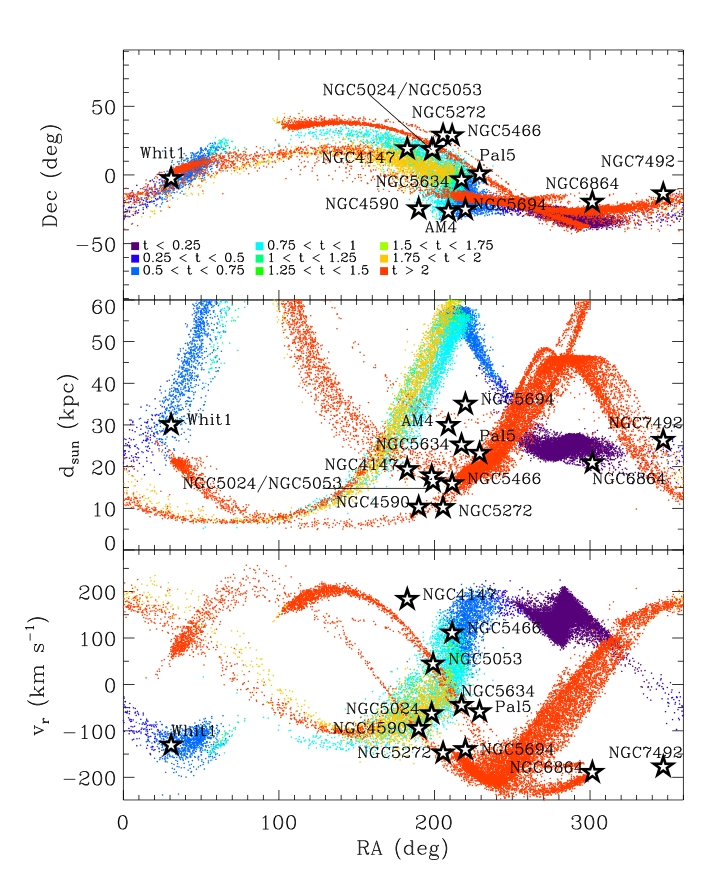}
       \caption[model]{{\small The Sgr tidal stream as presented in the model by P10. The upper panel shows the predicted orbit of the stream in the sky where the colour indicate different accretion times for the particles in ranges of 0.25\,Gyr long. The middle and bottom panels show the heliocentric distance and radial velocity distribution of the stream, using the same colour scheme. The position and radial velocity of the globulars in our sample are over-plotted as stars. Only the fraction of the substructure with $d_{\odot} <$ 60\,kpc has been considered. }}
\label{modelsagittarius}
     \end{center}
    \end{figure*}

   \begin{figure*}
     \begin{center}
      \includegraphics[scale=0.8]{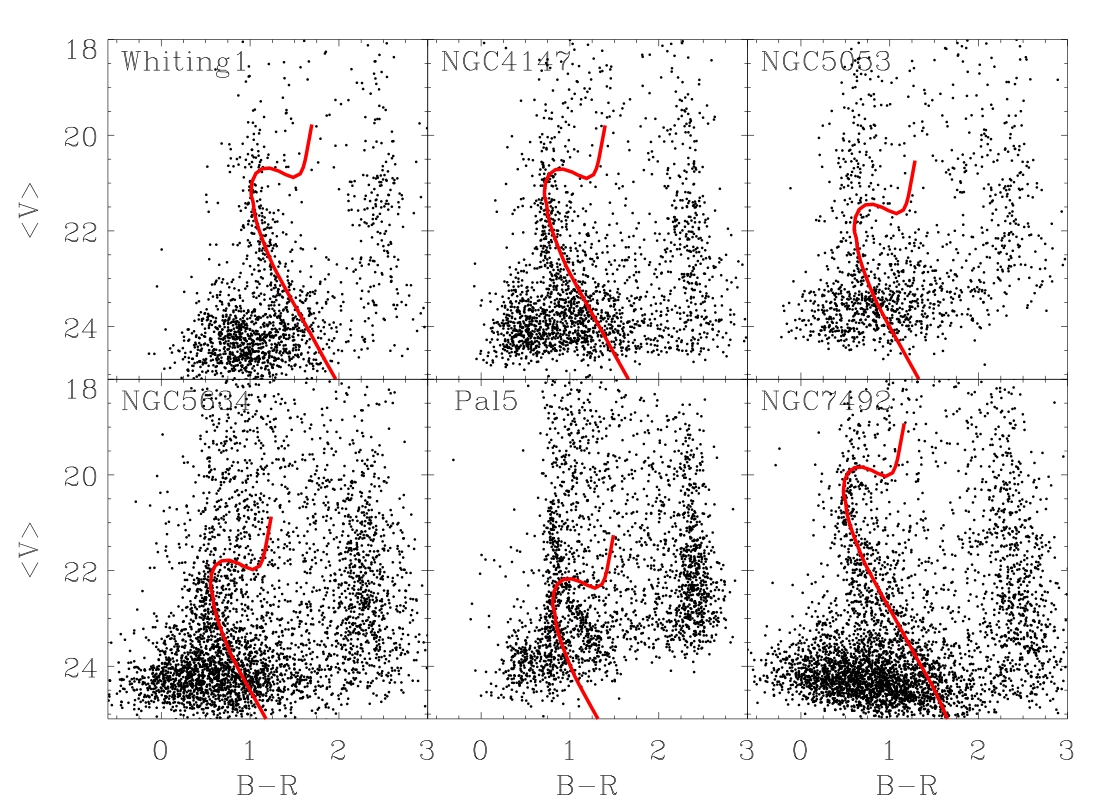}
       \caption[model]{{\small CMDs for the surrounding area of the GCs 
       possibly associated with the Sgr tidal stream, with stellar 
       over-densities that stand out in the comparison with the synthetic 
       diagrams. The Sgr isochrone corresponding to the radial distance derived 
       is over-plotted.}}
\label{resultados_sgr}
     \end{center}
    \end{figure*}

The stellar debris around clusters possibly associated with the Sgr tidal 
stream are, in general, the easiest cases to identify since the position and 
distance along the stream is well known from wide-sky surveys 
\citep{Majewski2003,Belokurov2006,Koposov2012} or numerous $N$-body simulations 
\citep[][hereafter P10]{Law2010a,Penarrubia2010}. To check the possible 
presence of Sgr tidal debris in our sample, we overplot the position and 
distances of our sample to the Sgr tidal stream model presented by P10 in the 
(RA, Dec) and $(\ell, d_{\odot})$ planes (\figref{modelsagittarius}). We find 
that 13 GCs of our sample lie within the projected position of the 
stream: Whiting\,1, NGC\,4147, NGC\,4590, NGC\,5024, NGC\,5053, NGC\,5272, 
AM\,4, NGC\,5466, NGC\,5634, NGC\,5694, Pal\,5, NGC\,6864 and NGC\,7492. 
In addition to this comparison with theoretical models, we compare the projected position of these clusters with 
the MS star density map of this structure from the SDSS by \cite{Koposov2012}. 
This shows that NGC\,5466 and NGC\,5272 are out of the projected 
path of the stream, which is consistent with our negative detections of tidal 
debris around these clusters. This could be also the case for NGC\,6864, 
NGC\,5694, NGC\,4590 and AM\,4 (see below).

Our survey around these Sgr stream GC candidates  reveals the clear presence of 
"probable" tidal debris from this stream around four of these
clusters (Whiting\,1, NGC\,4147, NGC\,5634 and Pal\,5; see CMDs in 
\figref{resultados_sgr}) plus a "possible" debris around NGC\,5053 and an
"uncertain" debris around NGC\,7492. 

One of the most significant ($S_{T},S_{B} > 10$) detections is 
unveiled in the area observed around the low-mass cluster Whiting\,1, which was 
also suggested as member of the Sgr GC system by \cite{Carraro2007}. The break 
in the radial stellar distribution found at $\sim$ 6\,arcmin 
\citep{Carraro2007} suggests that this low-density cluster is currently going 
through a disruption process due to the forces exerted by the Milky Way. 
However, it seems unlikely that the highly contrasted MS discovered in the area 
close to the cluster ($B-R \sim$ 1, $20.5 < V < 23$) shown in 
\figref{diagrams1}, lacking of any collimated spatial distribution, was 
generated by stars that have (or are close to) left  Whiting\,1. We identify 
the subjacent system as the trailing arm of the Sgr tidal strem and the 
position of the cluster relative to the stellar over-densities associated with 
that halo substructure supports that scenario \citep{Koposov2012}.

The isochrone fitting shows that Sgr and Whiting\,1 are spatially coincident, 
as also suggested by the cross-correlation results in 
\tabref{crosscorrelationresults}. The confirmation of the association of such a 
young GC \citep[6.5\,Gyr;][]{Carraro2007} will help to study  
the GC formation process in Sgr, given that Whiting\,1 might be the youngest GC among 
the clusters already associated with that dSph ($\sim 1$\,Gyr younger than the
intermediate--age GCs Arp\,2, Ter\,7 and Pal\,12, already associated to Sgr). This would indicate that Sgr 
was able to form GCs during a period of 6\,Gyr as pointed by 
\cite{Carraro2007}.

NGC\,5634 is one of the closest clusters to the plane that contains the orbit 
of the Sgr dSph (L10) and stream stars were identified by 
\cite{Majewski2003} in that line-of-sight. Our photometry shows for the first 
time a CMD morphology compatible with that of the Sgr stream in the surroundings of this cluster. 
It however does not reveal any underlying population at a similar 
distance of this cluster. A more important contribution in the background is 
detected, at a distance nearly twice the distance to NGC\,5634, as confirmed 
by both distance determination methods. On the basis of the P10 model, we identify 
that system in the background as a distant section of the leading arm of the 
Sgr tidal stream. 

The CMD of Pal\,5 presents the most complex morphology in our survey, 
displaying two MS-like features at different distances as shown in 
\figref{resultados_sgr}. The first lies is a high significance stellar 
population in the background of Pal\,5 at a similar distance of the 
cluster. These stars are likely cluster members populating the well-studied 
massive tidal tails emerging from this cluster \citep{Odenkirchen2001,Rockosi2002,Grillmair2006a}.
A second and significant ($S \sim 8$) MS is detected below the 
feature associated to the tidal tails \citep[see Fig.\,12 in ][]{Pila-Diez2014} 
at a radial distance compatible with that of the Sgr tidal stream according to 
P10. Interestingly, \cite{Sbordone2005} derived $\alpha$-element abundances for 
Pal\,5, resembling those obtained for M\,54 and Ter\,7, members of the Sgr GC 
system. 

\cite{Bellazzini2003} argued for the association of NGC 4147 with the Sgr 
stream from its radial velocity and the detection of M giant Sgr stars around 
this cluster. The detection of a MS feature from the Sgr stream stellar 
population around NGC\,4147 in our pencil-beam survey was already reported in 
\cite{MartinezDelgado2004}, before the mapping of this structure with 
large-scale surveys \citep[e.g. ][]{Majewski2003,Belokurov2006,Koposov2012}. 
We detect an underlying stellar population likely associated to that halo 
substructure at $d_{\odot} \sim$ 35\,kpc, separated from the GC along our 
line-of-sight by $\sim 15$\,kpc, in agreement with the position of the leading 
arm predicted by P10. Our results indicate that this cluster is not immersed in 
the Sgr tidal stream, as 
also pointed out in \cite{MartinezDelgado2004}, where the integrals of motions of 
both systems were analyzed. SDSS mapping has also showed that the path of the 
stream crosses the surroundings of NGC\,5024 and NGC\,5053, which are in the 
vicinity of NGC\,4147 in projected position \citep[see ][]{Koposov2012}. 

Around NGC\,5053 (classified in the group C) we have found an over-density in its background 
CMD suggesting a subjacent population at $\sim$ 40\,kpc, compatible with 
the radial distances predicted by P10 for the Sgr leading arms on that direction of the sky. The significance for the over-density 
in NGC\,5024 is $S < 5$. However, the cross-correlation method returns 
unconclusive or ambiguous detections in the case of both globulars, likely 
produced by the presence of blended populations in the diagrams, with a second 
MS possibly corresponding to a more distant wrap of Sgr with a low $S/N$. From 
these globulars, only NGC\,5053 is presented by L10 as a genuine candidate to 
belong to the Sgr GC system.

NGC\,7492 is the only cluster of our sample for which an "uncertain" detection of
an underlying debris (group B) has been found, and it is one of the globulars with low probability of 
belonging to the Sgr GC system according to L10. We identified a subjacent MS feature at a 
distance compatible with that of the cluster, which is not predicted in the synthetic 
TRILEGAL CMDs, while the significance of such a feature drops below the
adopted treshold when the Besan\c con model is adopted. With our photometry 
only, it is not possible to address the question of whether this detection is real or associated to
tidal tails originating from the cluster. However, in the radial profile obtained for this 
cluster \citep[see ][]{Carballo-Bello2012}, the stellar density beyond $r_{\rm bg}$ 
($\sim 14$\,arcmin) suggests the presence of a homogeneously distributed 
population. This suggests that the eventual underlying population is associated to 
a different system in the background of NGC\,7492. 
\figref{modelsagittarius} shows that the projected position and distance 
of the most recent accreted fraction of the Sgr stream trailing arm 
($t_{\rm accr} < 0.25$\,Gyr) is compatible with the position of this globular. 
Interestingly, the region around this cluster falls in a sky area without
SDSS data (see \figref{density6229_7006_7078}), but with evidence of Sgr stars 
in its vicinity, which strenghtens the hypothesis that this GC is embedded in 
the Sgr stream.

The negative detections in the surroundings of the other candidates prevent us 
from obtaining a final conclusion about the possible association of those 
clusters with the Sgr tidal stream, within our surface-brightness detection 
limits. Among them, only AM\,4 has been suggested as member of the Sgr GC 
system by \cite{Carraro2009} but, according to the background CMD obtained, 
there is no evidence of a subjacent stellar population associated to 
that stream. These negative detections, even in cases where the projected positions are favouring the 
detection of Sgr stream stars spatially coincident with the globulars (e.g. 
NGC\,5053 or NGC\,5634), might be used to establish the limitations of our 
photometric survey. Indeed, the absence of tidal remnants might be 
related to the evolution of the Sgr dSph and its interaction with the Milky Way. 
According to the model of P10, while Whiting\,1 and NGC\,7492 
are spatially coincident with the Sgr stars accreted in the last 0.75\,Gyr, 
NGC\,5053, NGC\,5634 and Pal\,5 are surrounded by
the material accreted from the satellite $>$ 2\,Gyr ago. This is a consequence of the fact that 
sections of the stream generated a long time ago are more dispersed, with a 
lower surface-brightness, and only the most recent arms of the Sgr tidal
stream could be detected by our survey . This scenario is also valid for 
Pal\,12, a cluster previously associated to Sgr by \cite{Martinez-Delgado2002}, 
which in the context of the P10 model seems to be associated with the section 
of the stream accreted in the last 0.75\,Gyr.   

In the bottom panel of \figref{modelsagittarius} we compare the predicted 
radial velocity of the stream with those values measured for the clusters in 
our sample \citep{Harris2010}. The globulars that are kinematically compatible 
and coincident with the position of the P10 tidal stream are Whiting\,1, 
NGC\,5053, NGC\,5634 and Pal\,5 (suggested as members of the Sgr GC system by 
L10). On the other hand, there is a difference of $\Delta v_{\rm r} \sim 100$\,km\,s$^{-1}$ in the case 
of NGC\,7492. So, for this stellar system, 
cluster and stream seem to be independent systems, although the orbit and 
structure of the Sgr stream in the southern sky are not well constrained 
because of the lack of a deep full-sky photometric database as the one available in the 
northern hemisphere (see discussion in L10, P10). Further follow-up spectroscopy is required 
to investigate the nature of the stellar population discovered around NGC\,7492. 

\subsection{Other over-densities}

The analysis of the CMDs corresponding to the GCs not associated with Sgr suggests the presence of MS features 
likely associated with subjacent stellar populations in three of them:
NGC1261, NGC1851 and NGC7006. In this section, we discuss the possible origin of these tentative remnants and their possible 
association with other known over-densities or stellar streams already reported 
in the Milky Way.

\subsubsection{An extended stellar over-density around NGC\,1851?}
\label{Discusion18511904}

One of the most conspicuous over-density of our survey, not associated with the 
Sgr stream, was detected around NGC 1851, first discovered by \cite{Olszewski2009}, 
 who interpreted this featuare as an {\it extended halo } surrounding 
this cluster up to distances of 75\,arcmin ($\sim 6.5 r_{\rm t}$) from the 
cluster centre, and independently reported by \citep{Carballo-Bello2010}.

NGC\,1851 is one of the most interesting candidates in our sample because of 
its multiple stellar populations \citep{Milone2008,Han2009} and the 
well-studied star-to-star abundance variations 
\cite[e.g.][]{Milone2009,Zoccali2009,Carretta2010,Carretta2011,Campbell2012,
Carretta2012}, which suggest a scenario in which this cluster is the result of 
the merging of two previous GCs, formed in the nucleus of an accreted dwarf 
galaxy \citep{Carretta2010,Bekki2012}. This cluster is member of a group of GCs 
formed by NGC\,1851, NGC\,1904, NGC\,2298 and NGC\,2808, which seems to be 
confined in a \emph{sphere} with a radius of 6\,kpc. That spatial distribution 
resembles that of M\,54, Terzan\,7, Terzan\,8 and Arp2, globulars found in the 
main body of the Sgr dSph \citep{Bellazzini2004,Martin2004}. In addition, all 
4 clusters show extended HB morphologies in their CMDs, feature that has been 
suggested as an indicator of an extra-Galactic origin in GCs \citep{Lee2007}. 

\figref{diagrams1} shows the presence of the prominent MS population in 
the surroundings of this cluster, which is the same reported by 
\cite{Olszewski2009}. This feature is not predicted by the TRILEGAL or 
Besan\c con models and it is also detected when the cross-correlation method is 
used (\tabref{crosscorrelationresults}) at a similar heliocentric distance than 
the cluster. Using low resolution spectra for a sample of 107 stars 
selected from the same photometry presented in this work, \cite{Sollima2012} detected a unexpected distinct 
stellar component with a radial velocity distribution that cannot be 
associated neither with the Galactic velocity field nor NGC\,1851 outliers, 
with a mean difference with respect to those components of $\Delta v_{\rm r} \sim 150$\, km\,s$^{-1}$ and 
$\Delta v_{\rm r} \sim 200$\, km\,s$^{-1}$, respectively. These authors
discuss the possible association of this feature with the Monoceros ring,
showing that the observed velocity distribution and the prediction 
made by the P05 model for that ring-like structure are slightly different,
although not completely inconsistent given the uncertainties in the adopted
Galactic potential. However, a recent spectroscopic analysis by \cite{Marino2014}
analysed a set of medium-resolution spectra for a sample of stars in the outer
halo of NGC\,1851 reporting the lack of any significant over-density of stars at
the velocity of such supposed stream. Summarizing, with the present dataset it 
is not clear if the detected over-density is linked to the precence of an
extended halo of cluster member stars \cite[as suggested by ][]{Olszewski2009} or
to a subjacent stream (possibly the Monoceros ring). Deep data extending over a wider FOV are
needed to distinguish between these two hypotheses.

\subsubsection{NGC\,7006}
\label{clustersheraq}

NGC\,7006 is a cluster slightly younger than other similar clusters 
in the inner Galaxy \citep{Dotter2011}. In addition, this GC is one of the most 
energetic clusters in the Milky Way with a very eccentric orbit 
\citep{Dinescu2001}, suggesting and extra-Galactic origin for that system. \figref{diagrams6} show the presence of a significant MS feature 
in the outer region of NGC\,7006 \citep[first reported in ][]{MartinezDelgado2004}. Since our cross-correlation method fails to detect these 
features due to the crowding of the fields (this cluster is classified in
the group C), our distance estimates are only based on isochrone fitting (\tabref{finalresults}). Our results 
show that the hypothetic subjacent stellar population is at different distance from the 
cluster. In particular, we derived a difference in distance of $\sim$ 8\,kpc for this possible tidal debris from the main body of
NGC\,7006. However, an inspection of the CMD of this cluster (\figref{diagrams6}) shows that the MS TO of this feature is severely 
affected by the presence of bright Milky Way disc stars at $V\sim$ 20--21, making the estimate of its position very uncertain. Therefore, 
we believe that this population lies at a distance $d_{\odot} = 15-20$\,kpc. 

 \begin{figure}
     \begin{center}
      \includegraphics[scale=0.4]{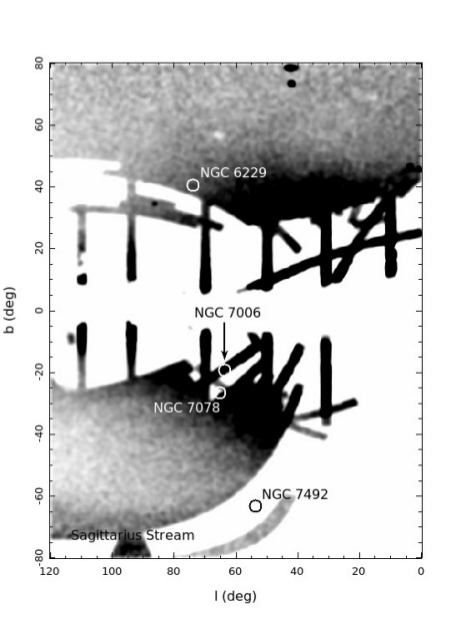}
       \caption[ngc6229]{{\small Density maps generated from SDSS data of the sky area where the GCs NGC\,6229, NGC\,7006, NGC\,7078 and NGC\,7492 are located. The huge stellar over-density observed might be associated whether with the Hercules-Aquila cloud \citep{Belokurov2007a,Simion2014} or with the region of higher density of halo stars reported by \cite{Deason2014}. Note the presence of the Sgr stream in the bottom left corner of the map.}}
\label{density6229_7006_7078}
     \end{center}
    \end{figure}

\figref{density6229_7006_7078} shows a stellar density map of MS-stars in a 
region of the sky from the SDSS photometric database, which includes both 
NGC7006 and NGC7078 (marked as open circles). These globulars seem to be 
inmersed in a region of high density of halo stars, that extends up to 
Galactic latitudes $b \sim -40^{\circ}$ \citep[see also ][]{Deason2014} and 
that might be the best explanation for the presence of this feature in the CMD
of NGC7006. However, an accurate model for the shape of the stellar halo is 
needed to confirm this possibility. 
 
An alternative scenario might be the presence of the southern component of the 
Hercules--Aquila over-density in the positions of this cluster. 
Recent results by \cite{Simion2014} supports the presence of a prominent 
over-density of RR\,Lyrae stars associated to this vast over-density in this 
region of the sky, with a distance range of $10 < d_{\odot} < 25$\,kpc (see 
their Fig. 9), strengthing the hypothesis of its origin from the tidal 
disruption of an ancient dwarf galaxy. That distance range is compatible with 
the one derived from our CMDs and suggests that NGC\,7006 might be well 
embedded in (and possibly associated to) this giant cloud of debris. 
Interestingly, \cite{Simion2014} also found that the Hercules--Aquila cloud is 
barely visible as a RR\, Lyrae over-density in the northern hemisphere, 
suggesting that this cloud is possibly not symmetric with respect to the 
Galactic plane. This is consistent with the low significance overdensity 
($S < 2.5$) of this component in the surroundings of NGC\,6229 (see 
\figref{diagrams5}).

\subsubsection{NGC\,1261}

NGC\,1261, lies in a projected position aligned with two other massive GCs 
showing an extended-HB morphology in their CMDs, NGC\,1851 and NGC\,1904. 
Around this cluster we have unveiled a stellar population (see 
\figref{diagrams1}) that stands out significantly when the background 
diagram is compared with the ones generated with the considered Galactic models and it 
is also apparent in the results obtained through the cross-correlation method 
(see \tabref{crosscorrelationresults}). The radial distance to the underlying 
component is similar to that of the cluster, suggesting that either it is 
composed of cluster members or of an unknown stellar population. The possible 
relation with the group of clusters described in \secref{Discusion18511904} 
encourages to explore the area between NGC\,1261 and those GCs.

\subsection{Negative detections}

There are no signatures of the presence of significant subjacent populations 
around the remaining candidates (AM4, NGC1904, NGC\,2298, NGC4590, NGC5024, 
NGC5272, NGC5466, NGC\,5694, NGC\,5824, NGC6229, NGC\,6864, 
NGC\,7078, Pal\,15 and Rup\,106) as we find no evidence 
of distinct stellar population concentrated at a specific distance within the 
probed colour-magnitude range using both the cross-correlation and the 
isochrone fitting methods. The photometric non-detection of tidal debris 
around the halo GCs in this study is an important result to consider in the 
context of hierarchical stellar halo assembly theories. Whether or not such 
non-detections can rule out an accretion origin for these GCs (and a portion of 
the Milky Way stellar halo) depends on two main factors: 1) how massive were 
the progenitor dwarf galaxies these GCs were accreted with, and 2) 
when were these dwarf galaxies and their GCs accreted into the Milky Way? 
Indeed, GCs hosted in low luminosity dwarfs which were accreted 
early, may show minimal associated stellar debris when observed at present.

\section{Conclusions}

We have presented the wide-field photometry of 23 Galactic GCs in the 
Galactocentric distance range 10 $\leq R_{\rm G} \leq$ 40\,kpc, searching in 
their surroundings for the stellar remnants of their accreted dwarf galaxy 
progenitors. We have detected a subjacent 
stellar populations beyond 1.5 times the $r_{\rm t}$ from the centre of 6 out 
the 23 GCs in our sample, and for three other clusters we found hints of
possible debris. These populations are in some cases consistent with known streams in the same 
line-of-sight of the GCs, while in other cases these over-densities might be 
associated with extended halos or tidal tails. Unfortunately our data do
not cover a region wide enough to detect the full extension of the observed
overdensities and their symmetry with respect to the cluster centre.

We identify the Sgr tidal stream in the direction of 6 GCs in our sample
(4 "probable", 1 "possible" and 1 "uncertain" detection )
and at distances compatible with the P10 orbital model. However,
the heliocentric distances to the subjacent populations are consistent with 
those of the related GCs only for 2 of them (Whiting\,1 
and NGC\,7492). Around NGC4147, NGC\,5634 and Pal\,5 (and with a smaller
level of significance NGC\,5053), previously suggested as 
members of the Sgr GC system, there are no significant detections
corresponding to the same cluster distance, although the signature of the Sgr 
MS is visible as a background feature. These negative detections might be 
related with our ability of unveiling faint subjacent tidal streams
 (at $\mu_{V}>32$\,mag\,sec$^{-2}$). It is possible that 
these globulars were accreted from the Sgr dSph a long time ago and 
the surface-brightness of the tidal remnants lies beyond our detection 
threshold above the fore/background populations. On the contrary, 
Whiting\,1 and NGC\,7492 seem to be immersed in the most recently accreted 
fraction of the stream ($<$ 0.75\,Gyr). Follow-up spectroscopy is needed to 
confirm the nature of the stellar population revealed by our photometry, more 
importantly in the latter case, where the detection is more uncertain and 
there exists a significant deviation between the radial velocities of the cluster and the prediction of the 
model by P10.

A subjacent stellar population has been also unveiled in the surroundings of 
NGC\,1851, NGC1261 and (possibly) NGC7006. These clusters lie far from the Sgr
predicted orbit and could be therefore related to other streams like the
Monoceros ring (NGC1851) the Hercules-Aquila cloud (NGC7006) or other unknown
debris.

\section*{Acknowledgements}

Based on observations made with the Isaac Newton Telescope operated on the island of La Palma by the Isaac Newton Group in the Spanish Observatorio del Roque de los Muchachos of the Instituto de Astrof\'isica de Canarias and with 2.2\,m ESO telescope at the La Silla Observatory under programme IDs 072.A-9002(A), 082.B-0386, 084.B-0666 and 085.B-0765. JC-B received partial support from Centre of Excellence in Astrophysics and Associated Technologies (PFB 06). AS acknowledges the PRIN MIUR 2010-2011 "The Chemical and Dynamical Evolution of the Milky Way and Local Group Galaxies" (PI. Matteucci). RL acknowledges financial support to the DAGAL network from the People Programme (Marie Curie Actions) of the European Unions Seventh Framework Programme FP7/2007- 2013/ under REA grant agreement number PITN-GA-2011-289313. R.~R.~M.~acknowledges partial support from CONICYT Anillo project ACT-1122 and project BASAL PFB-$06$ as well as FONDECYT project N$^{\circ}1120013$. JMC-S acknowledges financial support to CONICYT through the FONDECYT Postdoctoral Fellowship N$^{\circ} 3140310$. We warmly thank the anonymous referee for his/her helpful comments and suggestions. We thank J. Pe\~narrubia for allowing us to use the model of the Sagittarius tidal stream for this work. Thanks to A. Robin for her useful comments about the Besan\c con model. Thanks to A. Aparicio, S. Hidalgo, R. Carrera and D. C. Bardalez-Gagliuffi for their participation in the observing runs.

\def\jnl@style{\it}                       
\def\mnref@jnl#1{{\jnl@style#1}}          
\def\aj{\mnref@jnl{AJ}}                   
\def\apj{\mnref@jnl{ApJ}}                 
\def\apjl{\mnref@jnl{ApJL}}               
\def\aap{\mnref@jnl{A\&A}}                
\def\mnras{\mnref@jnl{MNRAS}}             
\def\nat{\mnref@jnl{Nat.}}                
\def\iaucirc{\mnref@jnl{IAU~Circ.}}       
\def\atel{\mnref@jnl{ATel}}               
\def\iausymp{\mnref@jnl{IAU~Symp.}}       
\def\pasp{\mnref@jnl{PASP}}               
\def\araa{\mnref@jnl{ARA\&A}}             
\def\apjs{\mnref@jnl{ApJS}}               
\def\aapr{\mnref@jnl{A\&A Rev.}}          

\bibliographystyle{mn2e}
\bibliography{biblio}

\begin{thebibliography}{}

\bibitem[\protect\citeauthoryear{{Bekki} \& {Yong}}{{Bekki} \&
  {Yong}}{2012}]{Bekki2012}
{Bekki} K.,  {Yong} D.,  2012, \mnras, 419, 2063

\bibitem[\protect\citeauthoryear{{Bellazzini}, {Correnti}, {Ferraro}, {Monaco}
  \& {Montegriffo}}{{Bellazzini} et~al.}{2006}]{Bellazzini2006}
{Bellazzini} M.,  {Correnti} M.,  {Ferraro} F.~R.,  {Monaco} L.,
  {Montegriffo} P.,  2006, \aap, 446, L1

\bibitem[\protect\citeauthoryear{{Bellazzini}, {Ferraro} \&
  {Ibata}}{{Bellazzini} et~al.}{2002}]{Bellazzini2002}
{Bellazzini} M.,  {Ferraro} F.~R.,    {Ibata} R.,  2002, \aj, 124, 915

\bibitem[\protect\citeauthoryear{{Bellazzini}, {Ferraro} \&
  {Ibata}}{{Bellazzini} et~al.}{2003}]{Bellazzini2003}
{Bellazzini} M.,  {Ferraro} F.~R.,    {Ibata} R.,  2003, \aj, 125, 188

\bibitem[\protect\citeauthoryear{{Bellazzini}, {Ibata} \&
  {Ferraro}}{{Bellazzini} et~al.}{2004}]{Bellazzini2004}
{Bellazzini} M.,  {Ibata} R.,    {Ferraro} F.~R.,  2004, in {F.~Prada,
  D.~Martinez Delgado, \& T.~J.~Mahoney} ed., Satellites and Tidal Streams
  Vol.~327 of Astronomical Society of the Pacific Conference Series, {Globular
  Clusters in the Sgr Stream and Other Structures}.
pp 220--+

\bibitem[\protect\citeauthoryear{{Bellazzini}, {Ibata}, {Martin}, {Lewis},
  {Conn} \& {Irwin}}{{Bellazzini} et~al.}{2006}]{Bellazzini2006a}
{Bellazzini} M.,  {Ibata} R.,  {Martin} N.,  {Lewis} G.~F.,  {Conn} B.,
  {Irwin} M.~J.,  2006, \mnras, 366, 865

\bibitem[\protect\citeauthoryear{{Bellazzini et al.}}{{Bellazzini et
  al.}}{2008}]{Bellazzini2008}
{Bellazzini et al.} 2008, \aj, 136, 1147

\bibitem[\protect\citeauthoryear{{Belokurov et al.}}{{Belokurov et
  al.}}{2006}]{Belokurov2006}
{Belokurov et al.} 2006, \apjl, 642, L137

\bibitem[\protect\citeauthoryear{{Belokurov et al.}}{{Belokurov et
  al.}}{2007a}]{Belokurov2007}
{Belokurov et al.} 2007a, \apj, 658, 337

\bibitem[\protect\citeauthoryear{{Belokurov et al.}}{{Belokurov et
  al.}}{2007b}]{Belokurov2007a}
{Belokurov et al.} 2007b, \apjl, 657, L89

\bibitem[\protect\citeauthoryear{{Bonaca et al.}}{{Bonaca et
  al.}}{2012}]{Bonaca2012}
{Bonaca et al.} 2012, \aj, 143, 105

\bibitem[\protect\citeauthoryear{{Bonifacio}, {Sbordone}, {Marconi}, {Pasquini}
  \& {Hill}}{{Bonifacio} et~al.}{2004}]{Bonifacio2004}
{Bonifacio} P.,  {Sbordone} L.,  {Marconi} G.,  {Pasquini} L.,    {Hill} V.,
  2004, \aap, 414, 503

\bibitem[\protect\citeauthoryear{{Bullock} \& {Johnston}}{{Bullock} \&
  {Johnston}}{2005}]{Bullock2005}
{Bullock} J.~S.,  {Johnston} K.~V.,  2005, \apj, 635, 931

\bibitem[\protect\citeauthoryear{{Campbell et al.}}{{Campbell et
  al.}}{2012}]{Campbell2012}
{Campbell et al.} 2012, \apjl, 761, L2

\bibitem[\protect\citeauthoryear{{Carballo-Bello}, {Gieles}, {Sollima},
  {Koposov}, {Mart{\'{\i}}nez-Delgado} \& {Pe{\~n}arrubia}}{{Carballo-Bello}
  et~al.}{2012}]{Carballo-Bello2012}
{Carballo-Bello} J.~A.,  {Gieles} M.,  {Sollima} A.,  {Koposov} S.,
  {Mart{\'{\i}}nez-Delgado} D.,    {Pe{\~n}arrubia} J.,  2012, \mnras, 419, 14

\bibitem[\protect\citeauthoryear{{Carballo-Bello} \&
  {Mart{\'{\i}}nez-Delgado}}{{Carballo-Bello} \&
  {Mart{\'{\i}}nez-Delgado}}{2010}]{Carballo-Bello2010}
{Carballo-Bello} J.~A.,  {Mart{\'{\i}}nez-Delgado} D.,  2010, in {Diego} J.~M.,
   {Goicoechea} L.~J.,  {Gonz{\'a}lez-Serrano} J.~I.,   {Gorgas} J.,  eds,
  Highlights of Spanish Astrophysics V {Tidal Remnants Around the Galactic
  Globular Clusters NGC1851 and NGC1904}.
p.~383

\bibitem[\protect\citeauthoryear{{Carollo et al.}}{{Carollo et
  al.}}{2007}]{Carollo2007}
{Carollo et al.} 2007, \nat, 450, 1020

\bibitem[\protect\citeauthoryear{{Carraro}}{{Carraro}}{2009}]{Carraro2009}
{Carraro} G.,  2009, \aj, 137, 3809

\bibitem[\protect\citeauthoryear{{Carraro}, {Zinn} \& {Moni Bidin}}{{Carraro}
  et~al.}{2007}]{Carraro2007}
{Carraro} G.,  {Zinn} R.,    {Moni Bidin} C.,  2007, \aap, 466, 181

\bibitem[\protect\citeauthoryear{{Carretta}, {D'Orazi}, {Gratton} \&
  {Lucatello}}{{Carretta} et~al.}{2012}]{Carretta2012}
{Carretta} E.,  {D'Orazi} V.,  {Gratton} R.~G.,    {Lucatello} S.,  2012, \aap,
  543, A117

\bibitem[\protect\citeauthoryear{{Carretta}, {Lucatello}, {Gratton},
  {Bragaglia} \& {D'Orazi}}{{Carretta} et~al.}{2011}]{Carretta2011}
{Carretta} E.,  {Lucatello} S.,  {Gratton} R.~G.,  {Bragaglia} A.,    {D'Orazi}
  V.,  2011, \aap, 533, A69

\bibitem[\protect\citeauthoryear{{Carretta et al.}}{{Carretta et
  al.}}{2010}]{Carretta2010}
{Carretta et al.} 2010, \apjl, 722, L1

\bibitem[\protect\citeauthoryear{{Chonis} \& {Gaskell}}{{Chonis} \&
  {Gaskell}}{2008}]{Chonis2008}
{Chonis} T.~S.,  {Gaskell} C.~M.,  2008, \aj, 135, 264

\bibitem[\protect\citeauthoryear{{Conn}, {Lewis}, {Irwin}, {Ibata}, {Ferguson},
  {Tanvir} \& {Irwin}}{{Conn} et~al.}{2005}]{Conn2005}
{Conn} B.~C.,  {Lewis} G.~F.,  {Irwin} M.~J.,  {Ibata} R.~A.,  {Ferguson}
  A.~M.~N.,  {Tanvir} N.,    {Irwin} J.~M.,  2005, \mnras, 362, 475

\bibitem[\protect\citeauthoryear{{Conn et al.}}{{Conn et al.}}{2007}]{Conn2007}
{Conn et al.} 2007, \mnras, 376, 939

\bibitem[\protect\citeauthoryear{{Cooper}, {D'Souza}, {Kauffmann}, {Wang},
  {Boylan-Kolchin}, {Guo}, {Frenk} \& {White}}{{Cooper}
  et~al.}{2013}]{Cooper2013}
{Cooper} A.~P.,  {D'Souza} R.,  {Kauffmann} G.,  {Wang} J.,  {Boylan-Kolchin}
  M.,  {Guo} Q.,  {Frenk} C.~S.,    {White} S.~D.~M.,  2013, \mnras, 434, 3348

\bibitem[\protect\citeauthoryear{{Cooper et al.}}{{Cooper et
  al.}}{2010}]{Cooper2010}
{Cooper et al.} 2010, \mnras, 406, 744

\bibitem[\protect\citeauthoryear{{Crane}, {Majewski}, {Rocha-Pinto},
  {Frinchaboy}, {Skrutskie} \& {Law}}{{Crane} et~al.}{2003}]{Crane2003}
{Crane} J.~D.,  {Majewski} S.~R.,  {Rocha-Pinto} H.~J.,  {Frinchaboy} P.~M.,
  {Skrutskie} M.~F.,    {Law} D.~R.,  2003, \apjl, 594, L119

\bibitem[\protect\citeauthoryear{{Deason}, {Belokurov}, {Koposov} \&
  {Rockosi}}{{Deason} et~al.}{2014}]{Deason2014}
{Deason} A.~J.,  {Belokurov} V.,  {Koposov} S.~E.,    {Rockosi} C.~M.,  2014,
  \apj, 787, 30

\bibitem[\protect\citeauthoryear{{Dinescu}, {Majewski}, {Girard} \&
  {Cudworth}}{{Dinescu} et~al.}{2000}]{Dinescu2000}
{Dinescu} D.~I.,  {Majewski} S.~R.,  {Girard} T.~M.,    {Cudworth} K.~M.,
  2000, \aj, 120, 1892

\bibitem[\protect\citeauthoryear{{Dinescu}, {Majewski}, {Girard} \&
  {Cudworth}}{{Dinescu} et~al.}{2001}]{Dinescu2001}
{Dinescu} D.~I.,  {Majewski} S.~R.,  {Girard} T.~M.,    {Cudworth} K.~M.,
  2001, \aj, 122, 1916

\bibitem[\protect\citeauthoryear{{Dinescu}, {Mart{\'{\i}}nez-Delgado},
  {Girard}, {Pe{\~n}arrubia}, {Rix}, {Butler} \& {van Altena}}{{Dinescu}
  et~al.}{2005}]{Dinescu2005}
{Dinescu} D.~I.,  {Mart{\'{\i}}nez-Delgado} D.,  {Girard} T.~M.,
  {Pe{\~n}arrubia} J.,  {Rix} H.-W.,  {Butler} D.,    {van Altena} W.~F.,
  2005, \apjl, 631, L49

\bibitem[\protect\citeauthoryear{{Dotter}, {Chaboyer}, {Jevremovi{\'c}},
  {Kostov}, {Baron} \& {Ferguson}}{{Dotter} et~al.}{2008}]{Dotter2008}
{Dotter} A.,  {Chaboyer} B.,  {Jevremovi{\'c}} D.,  {Kostov} V.,  {Baron} E.,
   {Ferguson} J.~W.,  2008, \apjs, 178, 89

\bibitem[\protect\citeauthoryear{{Dotter}, {Sarajedini} \& {Anderson}}{{Dotter}
  et~al.}{2011}]{Dotter2011}
{Dotter} A.,  {Sarajedini} A.,    {Anderson} J.,  2011, \apj, 738, 74

\bibitem[\protect\citeauthoryear{{Duffau}, {Zinn}, {Vivas}, {Carraro},
  {M{\'e}ndez}, {Winnick} \& {Gallart}}{{Duffau} et~al.}{2006}]{Duffau2006}
{Duffau} S.,  {Zinn} R.,  {Vivas} A.~K.,  {Carraro} G.,  {M{\'e}ndez} R.~A.,
  {Winnick} R.,    {Gallart} C.,  2006, \apjl, 636, L97

\bibitem[\protect\citeauthoryear{{Fellhauer et al.}}{{Fellhauer et
  al.}}{2006}]{Fellhauer2006}
{Fellhauer et al.} 2006, \apj, 651, 167

\bibitem[\protect\citeauthoryear{{Font}, {McCarthy}, {Crain}, {Theuns},
  {Schaye}, {Wiersma} \& {Dalla Vecchia}}{{Font} et~al.}{2011}]{Font2011a}
{Font} A.~S.,  {McCarthy} I.~G.,  {Crain} R.~A.,  {Theuns} T.,  {Schaye} J.,
  {Wiersma} R.~P.~C.,    {Dalla Vecchia} C.,  2011, \mnras, 416, 2802

\bibitem[\protect\citeauthoryear{{Font et al.}}{{Font et al.}}{2011}]{Font2011}
{Font et al.} 2011, \mnras, 417, 1260

\bibitem[\protect\citeauthoryear{{Forbes} \& {Bridges}}{{Forbes} \&
  {Bridges}}{2010}]{Forbes2010}
{Forbes} D.~A.,  {Bridges} T.,  2010, \mnras, 404, 1203

\bibitem[\protect\citeauthoryear{{Forbes}, {Strader} \& {Brodie}}{{Forbes}
  et~al.}{2004}]{Forbes2004}
{Forbes} D.~A.,  {Strader} J.,    {Brodie} J.~P.,  2004, \aj, 127, 3394

\bibitem[\protect\citeauthoryear{{Frinchaboy}, {Majewski}, {Crane}, {Reid},
  {Rocha-Pinto}, {Phelps}, {Patterson} \& {Mu{\~n}oz}}{{Frinchaboy}
  et~al.}{2004}]{Frinchaboy2004}
{Frinchaboy} P.~M.,  {Majewski} S.~R.,  {Crane} J.~D.,  {Reid} I.~N.,
  {Rocha-Pinto} H.~J.,  {Phelps} R.~L.,  {Patterson} R.~J.,    {Mu{\~n}oz}
  R.~R.,  2004, \apjl, 602, L21

\bibitem[\protect\citeauthoryear{{Gao}, {Just} \& {Grebel}}{{Gao}
  et~al.}{2013}]{Gao2013}
{Gao} S.,  {Just} A.,    {Grebel} E.~K.,  2013, \aap, 549, A20

\bibitem[\protect\citeauthoryear{{Girardi}, {Groenewegen}, {Hatziminaoglou} \&
  {da Costa}}{{Girardi} et~al.}{2005}]{Girardi2005}
{Girardi} L.,  {Groenewegen} M.~A.~T.,  {Hatziminaoglou} E.,    {da Costa} L.,
  2005, \aap, 436, 895

\bibitem[\protect\citeauthoryear{{Girardi et al.}}{{Girardi et
  al.}}{2010}]{Girardi10}
{Girardi et al.} 2010, \apj, 724, 1030

\bibitem[\protect\citeauthoryear{{G{\'o}mez}, {Helmi}, {Cooper}, {Frenk},
  {Navarro} \& {White}}{{G{\'o}mez} et~al.}{2013}]{Gomez2013}
{G{\'o}mez} F.~A.,  {Helmi} A.,  {Cooper} A.~P.,  {Frenk} C.~S.,  {Navarro}
  J.~F.,    {White} S.~D.~M.,  2013, \mnras, 436, 3602

\bibitem[\protect\citeauthoryear{{Grillmair}}{{Grillmair}}{2006}]{Grillmair2006b}
{Grillmair} C.~J.,  2006, \apjl, 645, L37

\bibitem[\protect\citeauthoryear{{Grillmair} \& {Dionatos}}{{Grillmair} \&
  {Dionatos}}{2006}]{Grillmair2006a}
{Grillmair} C.~J.,  {Dionatos} O.,  2006, \apjl, 641, L37

\bibitem[\protect\citeauthoryear{{Hammersley} \&
  {L{\'o}pez-Corredoira}}{{Hammersley} \&
  {L{\'o}pez-Corredoira}}{2011}]{Hammersley2011}
{Hammersley} P.~L.,  {L{\'o}pez-Corredoira} M.,  2011, \aap, 527, A6+

\bibitem[\protect\citeauthoryear{{Han}, {Lee}, {Joo}, {Sohn}, {Yoon}, {Kim} \&
  {Lee}}{{Han} et~al.}{2009}]{Han2009}
{Han} S.-I.,  {Lee} Y.-W.,  {Joo} S.-J.,  {Sohn} S.~T.,  {Yoon} S.-J.,  {Kim}
  H.-S.,    {Lee} J.-W.,  2009, \apjl, 707, L190

\bibitem[\protect\citeauthoryear{{Harris}}{{Harris}}{1991}]{Harris1991}
{Harris} W.~E.,  1991, \aj, 102, 1348

\bibitem[\protect\citeauthoryear{{Harris}}{{Harris}}{2010}]{Harris2010}
{Harris} W.~E.,  2010, preprint (arXiv:1012.3224)

\bibitem[\protect\citeauthoryear{{Helmi}, {Navarro}, {Meza}, {Steinmetz} \&
  {Eke}}{{Helmi} et~al.}{2003}]{Helmi2003}
{Helmi} A.,  {Navarro} J.~F.,  {Meza} A.,  {Steinmetz} M.,    {Eke} V.~R.,
  2003, \apjl, 592, L25

\bibitem[\protect\citeauthoryear{{Ibata}, {Irwin}, {Lewis}, {Ferguson} \&
  {Tanvir}}{{Ibata} et~al.}{2001}]{Ibata2001a}
{Ibata} R.,  {Irwin} M.,  {Lewis} G.,  {Ferguson} A.~M.~N.,    {Tanvir} N.,
  2001, \nat, 412, 49

\bibitem[\protect\citeauthoryear{{Ibata}, {Martin}, {Irwin}, {Chapman},
  {Ferguson}, {Lewis} \& {McConnachie}}{{Ibata} et~al.}{2007}]{Ibata2007}
{Ibata} R.,  {Martin} N.~F.,  {Irwin} M.,  {Chapman} S.,  {Ferguson} A.~M.~N.,
  {Lewis} G.~F.,    {McConnachie} A.~W.,  2007, \apj, 671, 1591

\bibitem[\protect\citeauthoryear{{Ibata}, {Gilmore} \& {Irwin}}{{Ibata}
  et~al.}{1994}]{Ibata1994}
{Ibata} R.~A.,  {Gilmore} G.,    {Irwin} M.~J.,  1994, \nat, 370, 194

\bibitem[\protect\citeauthoryear{{Ibata}, {Wyse}, {Gilmore}, {Irwin} \&
  {Suntzeff}}{{Ibata} et~al.}{1997}]{Ibata1997}
{Ibata} R.~A.,  {Wyse} R.~F.~G.,  {Gilmore} G.,  {Irwin} M.~J.,    {Suntzeff}
  N.~B.,  1997, \aj, 113, 634

\bibitem[\protect\citeauthoryear{{Juri{\'c} et al.}}{{Juri{\'c} et
  al.}}{2008}]{Juric2008}
{Juri{\'c} et al.} 2008, \apj, 673, 864

\bibitem[\protect\citeauthoryear{{King}}{{King}}{1966}]{King1966}
{King} I.~R.,  1966, \aj, 71, 64

\bibitem[\protect\citeauthoryear{{Koposov et al.}}{{Koposov et
  al.}}{2012}]{Koposov2012}
{Koposov et al.} 2012, \apj, 750, 80

\bibitem[\protect\citeauthoryear{{Landolt}}{{Landolt}}{1992}]{Landolt1992}
{Landolt} A.~U.,  1992, \aj, 104, 340

\bibitem[\protect\citeauthoryear{{Law} \& {Majewski}}{{Law} \&
  {Majewski}}{2010a}]{Law2010}
{Law} D.~R.,  {Majewski} S.~R.,  2010a, \apj, 718, 1128

\bibitem[\protect\citeauthoryear{{Law} \& {Majewski}}{{Law} \&
  {Majewski}}{2010b}]{Law2010a}
{Law} D.~R.,  {Majewski} S.~R.,  2010b, \apj, 714, 229

\bibitem[\protect\citeauthoryear{{Leaman}, {VandenBerg} \& {Mendel}}{{Leaman}
  et~al.}{2013}]{Leaman2013}
{Leaman} R.,  {VandenBerg} D.~A.,    {Mendel} J.~T.,  2013, \mnras, 436, 122

\bibitem[\protect\citeauthoryear{{Lee}, {Gim} \& {Casetti-Dinescu}}{{Lee}
  et~al.}{2007}]{Lee2007}
{Lee} Y.-W.,  {Gim} H.~B.,    {Casetti-Dinescu} D.~I.,  2007, \apjl, 661, L49

\bibitem[\protect\citeauthoryear{{L{\'o}pez-Corredoira}}{{L{\'o}pez-Corredoira}}{2006}]{Lopez-Corredoira2006}
{L{\'o}pez-Corredoira} M.,  2006, \mnras, 369, 1911

\bibitem[\protect\citeauthoryear{{Mackey et al.}}{{Mackey et
  al.}}{2010}]{Mackey2010}
{Mackey et al.} 2010, \apjl, 717, L11

\bibitem[\protect\citeauthoryear{{Mackey et al.}}{{Mackey et
  al.}}{2013}]{Mackey2013}
{Mackey et al.} 2013, \mnras, 429, 281

\bibitem[\protect\citeauthoryear{{Majewski}, {Skrutskie}, {Weinberg} \&
  {Ostheimer}}{{Majewski} et~al.}{2003}]{Majewski2003}
{Majewski} S.~R.,  {Skrutskie} M.~F.,  {Weinberg} M.~D.,    {Ostheimer} J.~C.,
  2003, \apj, 599, 1082

\bibitem[\protect\citeauthoryear{{Marigo}, {Girardi}, {Bressan}, {Groenewegen},
  {Silva} \& {Granato}}{{Marigo} et~al.}{2008}]{Marigo2008}
{Marigo} P.,  {Girardi} L.,  {Bressan} A.,  {Groenewegen} M.~A.~T.,  {Silva}
  L.,    {Granato} G.~L.,  2008, \aap, 482, 883

\bibitem[\protect\citeauthoryear{{Mar{\'{\i}}n-Franch et
  al.}}{{Mar{\'{\i}}n-Franch et al.}}{2009}]{Marin-Franch2009}
{Mar{\'{\i}}n-Franch et al.} 2009, \apj, 694, 1498

\bibitem[\protect\citeauthoryear{{Marino et al.}}{{Marino et
  al.}}{2014}]{Marino2014}
{Marino et al.} 2014, preprint (arXiv:1406.0944)

\bibitem[\protect\citeauthoryear{{Martin}, {Ibata}, {Bellazzini}, {Irwin},
  {Lewis} \& {Dehnen}}{{Martin} et~al.}{2004}]{Martin2004}
{Martin} N.~F.,  {Ibata} R.~A.,  {Bellazzini} M.,  {Irwin} M.~J.,  {Lewis}
  G.~F.,    {Dehnen} W.,  2004, \mnras, 348, 12

\bibitem[\protect\citeauthoryear{{Mart{\'{\i}}nez-Delgado}, {Aparicio},
  {G{\'o}mez-Flechoso} \& {Carrera}}{{Mart{\'{\i}}nez-Delgado}
  et~al.}{2001}]{Martinez-Delgado2001}
{Mart{\'{\i}}nez-Delgado} D.,  {Aparicio} A.,  {G{\'o}mez-Flechoso} M.~{\'A}.,
    {Carrera} R.,  2001, \apjl, 549, L199

\bibitem[\protect\citeauthoryear{{Mart{\'{\i}}nez-Delgado}, {Butler}, {Rix},
  {Franco}, {Pe{\~n}arrubia}, {Alfaro} \& {Dinescu}}{{Mart{\'{\i}}nez-Delgado}
  et~al.}{2005}]{Martinez-Delgado2005}
{Mart{\'{\i}}nez-Delgado} D.,  {Butler} D.~J.,  {Rix} H.-W.,  {Franco} V.~I.,
  {Pe{\~n}arrubia} J.,  {Alfaro} E.~J.,    {Dinescu} D.~I.,  2005, \apj, 633,
  205

\bibitem[\protect\citeauthoryear{{Mart{\'{\i}}nez-Delgado}, {Dinescu}, {Zinn},
  {Tutsoff}, {C{\^o}t{\'e}} \& {Boyarchuck}}{{Mart{\'{\i}}nez-Delgado}
  et~al.}{2004}]{MartinezDelgado2004}
{Mart{\'{\i}}nez-Delgado} D.,  {Dinescu} D.~I.,  {Zinn} R.,  {Tutsoff} A.,
  {C{\^o}t{\'e}} P.,    {Boyarchuck} A.,  2004, in {F.~Prada, D.~Martinez
  Delgado, \& T.~J.~Mahoney} ed., Satellites and Tidal Streams Vol.~327 of
  Astronomical Society of the Pacific Conference Series, {Mapping Tidal Streams
  around Galactic Globular Clusters}.
pp 255--+

\bibitem[\protect\citeauthoryear{{Mart{\'{\i}}nez-Delgado}, {Pe{\~n}arrubia},
  {Gabany}, {Trujillo}, {Majewski} \& {Pohlen}}{{Mart{\'{\i}}nez-Delgado}
  et~al.}{2008}]{Martinez-Delgado2008}
{Mart{\'{\i}}nez-Delgado} D.,  {Pe{\~n}arrubia} J.,  {Gabany} R.~J.,
  {Trujillo} I.,  {Majewski} S.~R.,    {Pohlen} M.,  2008, \apj, 689, 184

\bibitem[\protect\citeauthoryear{{Mart{\'{\i}}nez-Delgado}, {Pe{\~n}arrubia},
  {Juri{\'c}}, {Alfaro} \& {Ivezi{\'c}}}{{Mart{\'{\i}}nez-Delgado}
  et~al.}{2007}]{Martinez-Delgado2007}
{Mart{\'{\i}}nez-Delgado} D.,  {Pe{\~n}arrubia} J.,  {Juri{\'c}} M.,  {Alfaro}
  E.~J.,    {Ivezi{\'c}} Z.,  2007, \apj, 660, 1264

\bibitem[\protect\citeauthoryear{{Mart{\'{\i}}nez-Delgado}, {Zinn}, {Carrera}
  \& {Gallart}}{{Mart{\'{\i}}nez-Delgado} et~al.}{2002}]{Martinez-Delgado2002}
{Mart{\'{\i}}nez-Delgado} D.,  {Zinn} R.,  {Carrera} R.,    {Gallart} C.,
  2002, \apjl, 573, L19

\bibitem[\protect\citeauthoryear{{Mart{\'{\i}}nez-Delgado et
  al.}}{{Mart{\'{\i}}nez-Delgado et al.}}{2010}]{Martinez-Delgado2010}
{Mart{\'{\i}}nez-Delgado et al.} 2010, \aj, 140, 962

\bibitem[\protect\citeauthoryear{{Mateu}, {Vivas}, {Zinn}, {Miller} \&
  {Abad}}{{Mateu} et~al.}{2009}]{Mateu2009}
{Mateu} C.,  {Vivas} A.~K.,  {Zinn} R.,  {Miller} L.~R.,    {Abad} C.,  2009,
  \aj, 137, 4412

\bibitem[\protect\citeauthoryear{{McConnachie et al.}}{{McConnachie et
  al.}}{2009}]{McConnachie2009}
{McConnachie et al.} 2009, \nat, 461, 66

\bibitem[\protect\citeauthoryear{{McLaughlin} \& {van der Marel}}{{McLaughlin}
  \& {van der Marel}}{2005}]{McLaughlin2005}
{McLaughlin} D.~E.,  {van der Marel} R.~P.,  2005, \apjs, 161, 304

\bibitem[\protect\citeauthoryear{{Milone}, {Stetson}, {Piotto}, {Bedin},
  {Anderson}, {Cassisi} \& {Salaris}}{{Milone} et~al.}{2009}]{Milone2009}
{Milone} A.~P.,  {Stetson} P.~B.,  {Piotto} G.,  {Bedin} L.~R.,  {Anderson} J.,
   {Cassisi} S.,    {Salaris} M.,  2009, \aap, 503, 755

\bibitem[\protect\citeauthoryear{{Milone et al.}}{{Milone et
  al.}}{2008}]{Milone2008}
{Milone et al.} 2008, \apj, 673, 241

\bibitem[\protect\citeauthoryear{{Moitinho}, {V{\'a}zquez}, {Carraro}, {Baume},
  {Giorgi} \& {Lyra}}{{Moitinho} et~al.}{2006}]{Moitinho2006}
{Moitinho} A.,  {V{\'a}zquez} R.~A.,  {Carraro} G.,  {Baume} G.,  {Giorgi}
  E.~E.,    {Lyra} W.,  2006, \mnras, 368, L77

\bibitem[\protect\citeauthoryear{{Momany}, {Zaggia}, {Gilmore}, {Piotto},
  {Carraro}, {Bedin} \& {de Angeli}}{{Momany} et~al.}{2006}]{Momany2006}
{Momany} Y.,  {Zaggia} S.,  {Gilmore} G.,  {Piotto} G.,  {Carraro} G.,  {Bedin}
  L.~R.,    {de Angeli} F.,  2006, \aap, 451, 515

\bibitem[\protect\citeauthoryear{{Momany}, {Zaggia}, {Bonifacio}, {Piotto}, {De
  Angeli}, {Bedin} \& {Carraro}}{{Momany} et~al.}{2004}]{Momany2004}
{Momany} Y.,  {Zaggia} S.~R.,  {Bonifacio} P.,  {Piotto} G.,  {De Angeli} F.,
  {Bedin} L.~R.,    {Carraro} G.,  2004, \aap, 421, L29

\bibitem[\protect\citeauthoryear{{Newberg}, {Willett}, {Yanny} \&
  {Xu}}{{Newberg} et~al.}{2010}]{Newberg2010}
{Newberg} H.~J.,  {Willett} B.~A.,  {Yanny} B.,    {Xu} Y.,  2010, \apj, 711,
  32

\bibitem[\protect\citeauthoryear{{Newberg}, {Yanny} \& {Willett}}{{Newberg}
  et~al.}{2009}]{Newberg2009}
{Newberg} H.~J.,  {Yanny} B.,    {Willett} B.~A.,  2009, \apjl, 700, L61

\bibitem[\protect\citeauthoryear{{Newberg et al.}}{{Newberg et
  al.}}{2002}]{Newberg2002}
{Newberg et al.} 2002, \apj, 569, 245

\bibitem[\protect\citeauthoryear{{Odenkirchen et al.}}{{Odenkirchen et
  al.}}{2001}]{Odenkirchen2001}
{Odenkirchen et al.} 2001, \apjl, 548, L165

\bibitem[\protect\citeauthoryear{{Odenkirchen et al.}}{{Odenkirchen et
  al.}}{2003}]{Odenkirchen2003}
{Odenkirchen et al.} 2003, \aj, 126, 2385

\bibitem[\protect\citeauthoryear{{Olszewski}, {Saha}, {Knezek}, {Subramaniam},
  {de Boer} \& {Seitzer}}{{Olszewski} et~al.}{2009}]{Olszewski2009}
{Olszewski} E.~W.,  {Saha} A.,  {Knezek} P.,  {Subramaniam} A.,  {de Boer} T.,
    {Seitzer} P.,  2009, \aj, 138, 1570

\bibitem[\protect\citeauthoryear{{Palma}, {Majewski} \& {Johnston}}{{Palma}
  et~al.}{2002}]{Palma2002}
{Palma} C.,  {Majewski} S.~R.,    {Johnston} K.~V.,  2002, \apj, 564, 736

\bibitem[\protect\citeauthoryear{{Pe{\~n}arrubia}, {Belokurov}, {Evans},
  {Mart{\'{\i}}nez-Delgado}, {Gilmore}, {Irwin}, {Niederste-Ostholt} \&
  {Zucker}}{{Pe{\~n}arrubia} et~al.}{2010}]{Penarrubia2010}
{Pe{\~n}arrubia} J.,  {Belokurov} V.,  {Evans} N.~W.,
  {Mart{\'{\i}}nez-Delgado} D.,  {Gilmore} G.,  {Irwin} M.,
  {Niederste-Ostholt} M.,    {Zucker} D.~B.,  2010, \mnras, 408, L26

\bibitem[\protect\citeauthoryear{{Pe{\~n}arrubia et al.}}{{Pe{\~n}arrubia et
  al.}}{2005}]{Penarrubia2005}
{Pe{\~n}arrubia et al.} 2005, \apj, 626, 128

\bibitem[\protect\citeauthoryear{{Pe{\~n}arrubia et al.}}{{Pe{\~n}arrubia et
  al.}}{2011}]{Penarrubia2011}
{Pe{\~n}arrubia et al.} 2011, \apjl, 727, L2

\bibitem[\protect\citeauthoryear{{Peebles}}{{Peebles}}{1974}]{Peebles1974}
{Peebles} P.~J.~E.,  1974, \apjl, 189, L51+

\bibitem[\protect\citeauthoryear{{Pila-D{\'{\i}}ez}, {Kuijken}, {de Jong},
  {Hoekstra} \& {van der Burg}}{{Pila-D{\'{\i}}ez}
  et~al.}{2014}]{Pila-Diez2014}
{Pila-D{\'{\i}}ez} B.,  {Kuijken} K.,  {de Jong} J.~T.~A.,  {Hoekstra} H.,
  {van der Burg} R.~F.~J.,  2014, \aap, 564, A18

\bibitem[\protect\citeauthoryear{{Robin}, {Reyl{\'e}}, {Derri{\`e}re} \&
  {Picaud}}{{Robin} et~al.}{2003}]{Robin2003}
{Robin} A.~C.,  {Reyl{\'e}} C.,  {Derri{\`e}re} S.,    {Picaud} S.,  2003,
  \aap, 409, 523

\bibitem[\protect\citeauthoryear{{Rocha-Pinto}, {Majewski}, {Skrutskie} \&
  {Crane}}{{Rocha-Pinto} et~al.}{2003}]{Rocha-Pinto2003}
{Rocha-Pinto} H.~J.,  {Majewski} S.~R.,  {Skrutskie} M.~F.,    {Crane} J.~D.,
  2003, \apjl, 594, L115

\bibitem[\protect\citeauthoryear{{Rockosi et al.}}{{Rockosi et
  al.}}{2002}]{Rockosi2002}
{Rockosi et al.} 2002, \aj, 124, 349

\bibitem[\protect\citeauthoryear{{Sales et al.}}{{Sales et
  al.}}{2008}]{Sales2008}
{Sales et al.} 2008, \mnras, 389, 1391

\bibitem[\protect\citeauthoryear{{Sbordone}, {Bonifacio}, {Marconi}, {Buonanno}
  \& {Zaggia}}{{Sbordone} et~al.}{2005}]{Sbordone2005}
{Sbordone} L.,  {Bonifacio} P.,  {Marconi} G.,  {Buonanno} R.,    {Zaggia} S.,
  2005, \aap, 437, 905

\bibitem[\protect\citeauthoryear{{Schlafly} \& {Finkbeiner}}{{Schlafly} \&
  {Finkbeiner}}{2011}]{Schlafly2011}
{Schlafly} E.~F.,  {Finkbeiner} D.~P.,  2011, \apj, 737, 103

\bibitem[\protect\citeauthoryear{{Schlegel}, {Finkbeiner} \&
  {Davis}}{{Schlegel} et~al.}{1998}]{Schlegel1998}
{Schlegel} D.~J.,  {Finkbeiner} D.~P.,    {Davis} M.,  1998, \apj, 500, 525

\bibitem[\protect\citeauthoryear{{Searle} \& {Zinn}}{{Searle} \&
  {Zinn}}{1978}]{Searle1978}
{Searle} L.,  {Zinn} R.,  1978, \apj, 225, 357

\bibitem[\protect\citeauthoryear{{Siegel et al.}}{{Siegel et
  al.}}{2007}]{Siegel2007}
{Siegel et al.} 2007, \apjl, 667, L57

\bibitem[\protect\citeauthoryear{{Simion}, {Belokurov}, {Irwin} \&
  {Koposov}}{{Simion} et~al.}{2014}]{Simion2014}
{Simion} I.~T.,  {Belokurov} V.,  {Irwin} M.,    {Koposov} S.~E.,  2014,
  \mnras, 440, 161

\bibitem[\protect\citeauthoryear{{Skrutskie et al.}}{{Skrutskie et
  al.}}{2006}]{Skrutskie2006}
{Skrutskie et al.} 2006, \aj, 131, 1163

\bibitem[\protect\citeauthoryear{{Sollima}, {Gratton}, {Carballo-Bello},
  {Mart{\'{\i}}nez-Delgado}, {Carretta}, {Bragaglia}, {Lucatello} \&
  {Pe{\~n}arrubia}}{{Sollima} et~al.}{2012}]{Sollima2012}
{Sollima} A.,  {Gratton} R.~G.,  {Carballo-Bello} J.~A.,
  {Mart{\'{\i}}nez-Delgado} D.,  {Carretta} E.,  {Bragaglia} A.,  {Lucatello}
  S.,    {Pe{\~n}arrubia} J.,  2012, \mnras, 426, 1137

\bibitem[\protect\citeauthoryear{{Sollima}, {Valls-Gabaud}, {Martinez-Delgado},
  {Fliri}, {Pe{\~n}arrubia} \& {Hoekstra}}{{Sollima}
  et~al.}{2011}]{Sollima2011}
{Sollima} A.,  {Valls-Gabaud} D.,  {Martinez-Delgado} D.,  {Fliri} J.,
  {Pe{\~n}arrubia} J.,    {Hoekstra} H.,  2011, \apjl, 730, L6+

\bibitem[\protect\citeauthoryear{{Stetson}}{{Stetson}}{1987}]{Stetson1987}
{Stetson} P.~B.,  1987, \pasp, 99, 191

\bibitem[\protect\citeauthoryear{{Strader}, {Brodie}, {Forbes}, {Beasley} \&
  {Huchra}}{{Strader} et~al.}{2003}]{Strader2003}
{Strader} J.,  {Brodie} J.~P.,  {Forbes} D.~A.,  {Beasley} M.~A.,    {Huchra}
  J.~P.,  2003, \aj, 125, 1291

\bibitem[\protect\citeauthoryear{{Tonini}}{{Tonini}}{2013}]{Tonini2013}
{Tonini} C.,  2013, \apj, 762, 39

\bibitem[\protect\citeauthoryear{{Trager}, {King} \& {Djorgovski}}{{Trager}
  et~al.}{1995}]{Trager1995}
{Trager} S.~C.,  {King} I.~R.,    {Djorgovski} S.,  1995, \aj, 109, 218

\bibitem[\protect\citeauthoryear{{van den Bergh} \& {Mackey}}{{van den Bergh}
  \& {Mackey}}{2004}]{vanden2004}
{van den Bergh} S.,  {Mackey} A.~D.,  2004, \mnras, 354, 713

\bibitem[\protect\citeauthoryear{{VandenBerg}, {Brogaard}, {Leaman} \&
  {Casagrande}}{{VandenBerg} et~al.}{2013}]{VandenBerg2013}
{VandenBerg} D.~A.,  {Brogaard} K.,  {Leaman} R.,    {Casagrande} L.,  2013,
  \apj, 775, 134

\bibitem[\protect\citeauthoryear{{Vanhollebeke}, {Groenewegen} \&
  {Girardi}}{{Vanhollebeke} et~al.}{2009}]{Vanhollebeke2009}
{Vanhollebeke} E.,  {Groenewegen} M.~A.~T.,    {Girardi} L.,  2009, \aap, 498,
  95

\bibitem[\protect\citeauthoryear{{Williams et al.}}{{Williams et
  al.}}{2011}]{Williams2011}
{Williams et al.} 2011, \apj, 728, 102

\bibitem[\protect\citeauthoryear{{Yanny et al.}}{{Yanny et
  al.}}{2003}]{Yanny2003}
{Yanny et al.} 2003, \apj, 588, 824

\bibitem[\protect\citeauthoryear{{York et al.}}{{York et al.}}{2000}]{York2000}
{York et al.} 2000, \aj, 120, 1579

\bibitem[\protect\citeauthoryear{{Zinn}}{{Zinn}}{1993}]{Zinn1993}
{Zinn} R.,  1993, in {G.~H.~Smith \& J.~P.~Brodie} ed., The Globular
  Cluster-Galaxy Connection Vol.~48 of Astronomical Society of the Pacific
  Conference Series, {The Galactic Halo Cluster Systems: Evidence for
  Accretion}.
pp 38--+

\bibitem[\protect\citeauthoryear{{Zoccali}, {Pancino}, {Catelan}, {Hempel},
  {Rejkuba} \& {Carrera}}{{Zoccali} et~al.}{2009}]{Zoccali2009}
{Zoccali} M.,  {Pancino} E.,  {Catelan} M.,  {Hempel} M.,  {Rejkuba} M.,
  {Carrera} R.,  2009, \apjl, 697, L22

\bibitem[\protect\citeauthoryear{{Zolotov}, {Willman}, {Brooks}, {Governato},
  {Brook}, {Hogg}, {Quinn} \& {Stinson}}{{Zolotov} et~al.}{2009}]{Zolotov2009}
{Zolotov} A.,  {Willman} B.,  {Brooks} A.~M.,  {Governato} F.,  {Brook} C.~B.,
  {Hogg} D.~W.,  {Quinn} T.,    {Stinson} G.,  2009, \apj, 702, 1058

\end{thebibliography}

\end{document}